\documentclass[journal=mamobx,manuscript=article]{achemso}
\usepackage[version=3]{mhchem} % Formula subscripts using \ce{}
\usepackage[T1]{fontenc}
\usepackage{graphicx}
\usepackage{dcolumn}
\usepackage{amsmath}
\usepackage{amssymb}
\usepackage{amsfonts}
\usepackage{bm}
\usepackage{color}
\usepackage{tikz}
\usepackage{atbegshi}
\usetikzlibrary{shapes,patterns,arrows}
\newcommand{\be}{\begin{equation}}
\newcommand{\ee}{\end{equation}}
\newcommand{\ba}{\begin{eqnarray}}
\newcommand{\ea}{\end{eqnarray}}
\setkeys{acs}{articletitle=true}
\definecolor{green4}{rgb}{0,0.55,0}

\author{Gianmarco Muna\`o$^1$}
\email{gmunao@unisa.it}
\author{Antonio De Nicola$^2$}
\author{Florian M{\"u}ller-Plathe$^3$}
\author{Toshihiro Kawakatsu$^4$}
\author{Andreas Kalogirou$^3$}
\author{Giuseppe Milano$^{1,2}$}
\affiliation{
$^{1}$Dipartimento di Chimica e Biologia, Universit\`a di Salerno,
Via Giovanni Paolo II, 132, I-84084, Fisciano (SA), Italy. \\
$^{2}$Department of Organic Materials Science, Yamagata University,
4-3-16 Jonan Yonezawa, Yamagata-ken 992-8510, Japan. \\
$^{3}$Eduard-Zintl-Institut f{\"u}r Anorganische und Physikalische Chemie and Center of Smart Interfaces, Technische Universit{\"a}t Darmstadt, Alarich-Weiss-Str. 8, 64287 Darmstadt, Germany. \\
$^{4}$Department of Physics, Tohoku University, Aoba, Aramaki, Aoba-ku, Sendai, Miyagi 980-8578, Japan. \\
}

\title{Influence of polymer bidispersity  
on the effective particle-particle interactions in polymer nanocomposites}


\begin{document}

\begin{abstract}
We investigate the role played by the bidispersity of polymer chains on 
the local structure and the potential of mean force (PMF) 
between silica nanoparticles
(NPs) in a polystyrene melt. We use the 
hybrid particle-field 
molecular dynamics technique which allows to efficiently relax 
polymer nanocomposites even with high molecular weights. 
The NPs we investigate are 
either bare or grafted with polystyrene chains immersed in a melt of
free polystyrene chains, whereas the grafted and the free polystyrene chains
are either monodisperse or bidisperse. 
The two-body PMF shows that a
bidisperse distribution of free 
polymer chains increases the strength of attraction
between a pair of ungrafted NPs. 
If the NPs are grafted by polymer chains, the effective
interaction crucially depends on bidispersity and grafting density of the 
polymer chains: 
for low
grafting densities, the bidispersity of both free and grafted 
chains increases the repulsion
between the NPs, whereas for high grafting densities 
we observe two different
effects. An increase of bidispersity in free chains causes
the rise of the repulsion
between the NPs, while an increase of bidispersity in grafted chains promotes
the rise of attraction.
Additionally, a proper treatment of 
multi-body interactions improves the simpler 
two-body PMF calculations, in both unimodal and bimodal cases.
We found that, by properly tuning the bidispersity of both
free and grafted chains, we can control the structure of the composite 
materials, which can be confirmed by experimental observations. 
As a results, the hybrid particle-field approach is confirmed to be
a valid tool for reproducing and predicting microscopic 
interactions, 
which determine the stability of the microscopic structure of
the composite in a wide range of conditions.
\end{abstract}

\section{Introduction}
A proper control of microscopic structural 
arrangements and interactions between components is crucial
in order to design a polymer nanocomposite material with specific 
macroscopic properties. In this context, it has been well 
documented~\cite{Akcora:09,Akcora:10,Kumar:17} that nanoparticles (NPs)
dispersed in a polymer matrix tend to aggregate.
Such a tendency depends on
the NP size, the molecular weight of polymer chains and the NP 
concentration. Since superior mechanical properties of the composite material 
can be obtained if the fillers are dispersed, rather than  
forming aggregates, the 
NPs are often grafted with polymer chains. 
In this case, the grafted polymer chains can be either the same chemical
species as the polymer in the matrix or different species.
On the other hand, experiments~\cite{Lan:07,Chevigny:10,Sunday:12,Archer:12} 
have largely proved that grafted NPs can be either well dispersed 
or arranged in nanoaggregates. The onset of a given morphology 
depends on both the ratio between the molecular weights of
free and grafted polymer chains and the grafting 
density~\cite{Kumar:13}. Since the interaction between the polymer chains
is essential to determine 
the phase behavior of the composite, a desired
morphology can be obtained also by properly tuning  
the polydispersity of the polymer chains.
In recent years, 
many 
experimental~\cite{Park:18,Rungta:12,Natarajan:13,Schadler:13,Triebel:11} 
as well as theoretical and simulation
works~\cite{Dodd:12,Martin:13,Martin:13a,Martin:13b,Martin:14, Langeloth:14,Sci:17} have been devoted to the investigation of 
the role played by the polydispersity 
in the
characterization of the phase behavior and the stability of nanocomposites. 
Experimental 
studies~\cite{Park:18,Rungta:12,Natarajan:13,Schadler:13,Triebel:11} 
have shown that if free and grafted polymer chains  
belong to the same chemical
species, NPs with a unimodal distribution of
grafted chains tend to aggregate, 
provided that the molecular weight of the grafted chains is lower than that 
of the free chains.
On the other hand, by mixing short and long grafted chains, the interparticle
attraction can be lowered considerably, 
generating a well dispersed condition.
More specifically, much interest is dedicated 
to low 
grafting densities, since in this case
complex self-assembled
nanostructures can be expected~\cite{Akcora:09,Akcora:10}.
A recent experimental work~\cite{Natarajan:13} on silica NPs 
in a polystyrene (PS) matrix, has shown
that for low grafting density, in comparison with the
monodisperse case, a proper choice of the 
polydispersity of the grafted chains can enlarge the stable region 
of the well-dispersed NP phase while makes
the self-assembled phase narrow. 
At the
same time, the authors showed that the polydispersity can be used 
to obtain a desired structure as well, 
whereas different aggregates appear for 
unimodal   
distributions of grafted chains. 
Simulation studies~\cite{Martin:13,Martin:13a,Martin:13b,Sci:17} 
have proved that the attractive or 
repulsive behavior of the effective interactions (and hence the onset of 
aggregates) is due to the wettability
of the NP surfaces with the free polymers, which, in turn, 
depends on the local structure of polymer chains.
An accurate description of the chain arrangements in proximity of the NP
surface and the interaction between free and grafted chains is crucial 
to predict 
the final behavior of the composite. However the understanding
of these mechanisms at a molecular level is not straightforward, 
due to the number of interactions that should be taken under control.
In order to overcome these issues, many simulation
studies of silica NPs in a PS melt have been performed. More specifically,
structure and dynamics in the melt state and elastic constants in the glassy 
state of systems consisting of bare silica NPs immersed in long-chain 
polystyrene matrices have been simulated in Ref.~\cite{Mathioudakis:16}.
Structure around a silica NP bearing surface grafted polystyrene chains in 
a molten polystyrene matrix has been explored in Ref.~\cite{Vogiatzis:13}
while a similar study for bare nanoparticles has been conducted in
Ref.~\cite{Vogiatzis:11a}. \\
Despite the large number of studies carried out so far, 
there are still questions that need to be addressed: 
from a simulation point
of view, it is worth pointing out that generally 
only non-specific bead-spring polymer models  
and hard spheres for NPs have been adopted, 
which disregard information on the chemical nature of
the nanocomposites.  
A further point concerns the importance of 
contributions to the total interaction from the 
three-body (or, in general,
multi-body) effects. This is particularly true in cases of polymer chains
with high molecular weight, where a single chain can easily interact with
more than two NPs, hence making the multi-body contributions not negligible,
as shown in recent simulations~\cite{Munao:18a} and
theoretical~\cite{Yeth:11} works. 

We propose a simulation study of the two-body and three-body 
potentials of mean force (PMF) 
among 2 or 3 silica NPs (bare or grafted with PS chains) 
embedded in a PS matrix. 
We adopt the hybrid particle-field molecular dynamics
approach~\cite{Milano:09} which has been recently employed 
to efficiently
characterize, among others, 
nanocomposites~\cite{Denicola:15,Denicola:16,Munao:18a}, 
polymer melts of high molecular 
weight~\cite{Denicola:14} and carbon nanotubes~\cite{Zhao:16}.
In addition, a number of studies have been devoted to capture 
structure, thermodynamics, and also entangled polymer dynamics of polymer 
melts in the bulk and at 
interfaces~\cite{Sgouros:17,Vogiatzis:17,Megariotis:18,Sgouros:18}.
The advantage to use a hybrid approach relies on the possibility
to significantly speed up the simulation time typically required 
to get a  
proper relaxion of polymer nanocomposites~\cite{Denicola:14}. This speed up 
is achieved by decoupling the mutual particle-particle interactions and
replacing them with a field representation. 
According to the hybrid approach the density fields are obtained by 
calculating ``on the fly'' the density values from the particle positions. 
This prescription allows us
to perform realistic simulations of systems with
a molecular/atomistic detail, with applications to 
polymers~\cite{Denicola:14,Denicola:15a}, biological models
like phospholipids~\cite{Denicola:11}, proteins~\cite{Bore:18}
and systems for drug release~\cite{Denicola:14a}. In addition,
this scheme paves the way to the
study of equilibrium properties (like the PMF), 
whose treatment with traditional simulation
approach is difficult~\cite{Munao:18a}.
In the present work the grafting density and the polydispersity of both free 
and grafted polymer chains are varied, and 
realistic coarse-grained (CG) models are needed, which have  
to account for the
chemical detail of the system. We consider two different
molecular weights of polymer chains, {\it i.e.} we use 
a bidisperse system.
In order to ascertain the role played by
free and grafted chains, we separately investigate the effects  
of the bidispersity of both of these two types of chains. 

%%%%%%%%%%%%%%%%%%%%%%%%%%%%%%%%%%%%%%%%%%%%%%%%%%%%%%
\section{Simulation protocol}
The models for silica NPs and for PS chains, together with a 
proposed 
simulation method, have been already adopted 
to successfully investigate
monodisperse systems~\cite{Denicola:16,Munao:18a}. All details concerning  
these models and the relative simulation strategies can be found in those
works and in references therein, although 
their general description is 
reported in the following subsections. 

\subsection{Coarse-grained models and hybrid particle-field approach}
The CG models adopted in the present work
have been developed by Qian and
coworkers~\cite{Qian:08} and successfully implemented in
Refs.~\cite{Muller-Plathe:12,Denicola:16,Munao:18a}. In the CG model of 
atactic PS,
a repeating unit is replaced by
a single bead placed at its center of mass. 
Bond length, angle, dihedral
and torsional parameters are set in order to reproduce the local 
structure of atomistic PS models. Also, two different kinds of
beads account for the chirality of the asymmetric carbon.
In the CG model of the NP
a single bead is centered on the silicon atom position
and represents one ${\rm SiO_2}$ unit. The total mass of the 
surface hydrogen atoms
is distributed over all the beads of the
NP. The simulated NP has a diameter of 4 nm and contains 873 beads.
In addition, we also consider systems where some PS chains are grafted to
the NP surface. Following the prescription by Ghanbari and
coworkers~\cite{Muller-Plathe:12}, each grafted chain is attached to the NP
surface through a linker unit, that is divided into four beads of two kinds
with the same mass. 
Further details on this CG representation and
its validation can be found in
Refs.~\cite{Muller-Plathe:12,Denicola:16}.
In those works it has been demonstrated that local structure, 
gyration radius and spatial orientation of free and grafted chains well 
reproduce the atomistic models studied in Refs.~\cite{Ndoro:11,Ndoro:12}. 
The latter, in turn, 
show a good agreement with experiments in describing the  structural 
properties of the polymer in the vicinity of the NP, concerning in particular 
the layering of the chains and the interfacial properties. In addition, a
further validation of our approach has been performed in 
Ref.~\cite{Munao:18a} in which the experimental phase behavior of 
silica-polystyrene nanocomposites has been qualitatively reproduced by a 
proper combination of two-body and three-body PMF.

The simulation approach that we adopt in the present work 
is based on a combination of a standard molecular
dynamics (MD) approach and a self-consistent field (SCF) 
theory~\cite{Kawakatsu:04}
for the calculation of non-bonded potentials. 
The resulting scheme is known as
hybrid particle-field model~\cite{Milano:09}: in this approach,
the Hamiltonian of a system of $M$ molecules is split as:
\begin{equation}\label{eq:H}
\hat{H}(\Gamma)=\hat{H_0}(\Gamma)+\hat{J}(\Gamma)
\end{equation}
where $\Gamma$ represents a point in the phase space and the symbol \hspace{0.01cm} $\hat{}$
\hspace{0.01cm} indicates that a given quantity is a function of the microscopic state
corresponding to $\Gamma$. 
In Eq.~\ref{eq:H}, $\hat{H_0}(\Gamma)$ is the
Hamiltonian of a system where the molecules experience 
only intramolecular interactions, whereas 
$\hat{J}(\Gamma)$ is the contribution due to all
non-bonded interactions.
The latter can be calculated as an external potential $V({\bf r})$ on single
particles, which is due to the density field. The details of the derivation
of $V({\bf r})$ can be found elsewhere~\cite{Milano:09}. 
The mean field solution for the
potential $V_K({\bf r})$ acting on a particle of type $K$ 
at position ${\bf r}$, is:
\begin{equation}\label{eq:pot}
V_K({\bf r})=k_B T \sum_{K'} \chi_{KK'} \Phi_{K'}({\bf r}) +
\frac{1}{\kappa}(\sum_{K'} \Phi_{K'} ({\bf r}) -1)
\end{equation}
where $k_B$ is the Boltzmann constant, $T$ is the temperature,
$\chi_{KK'}$ are the mean field parameters for the interaction of a
particle of type $K$ with the density field due to particles of type $K'$ and
the second term on the right-hand
side of Eq.~\ref{eq:pot} is the incompressibility condition,
$\kappa$ being the compressibility. Also, $\Phi_K ({\bf r})$ and $\Phi_{K'}({\bf r})$
are the density functions of the beads of types 
$K$ and $K'$, respectively, 
and normalized by the bulk density value $\phi_0$.
For the NP-PS interactions $\chi_{KK'} \times RT = 5.25$ kJ/mol 
($R$ being the gas constant), while for NP-NP and PS-PS interactions 
$\chi_{KK'} \times RT = 0$, in agreement with a
previous MD-SCF study of the same model~\cite{Denicola:16}.

All simulations have been performed by means of the OCCAM code, whose
details can be found in Ref.~\cite{Occam}, and run 
in the {\it NVT} ensemble, with the
temperature (fixed at 590 K) controlled by the Andersen thermostat and a
time step of 4 fs. 
The density fields, with the corresponding 
derivatives, are calculated on a lattice obtained by dividing the simulation 
box into a grid of cells, whose spacing has been set to 1.18 nm. 
The density is calculated 
from particle positions and projected on the mesh according to a 
procedure described in Ref.~\cite{Denicola:16}. The mesh is
updated every 100 MD time steps.
The initial configurations
have been built by using the Packmol program~\cite{Packmol}.
All particles have been
enclosed in a simulation box with
periodic boundary conditions: box sizes, along with the number
of particles simulated, are reported in 
Tabs.~\ref{tab:1NP} and~\ref{tab:2PMF}.
The total force acting on each particle is a 
sum of intramolecular
interactions (obtained by means of tabulated potentials)
and contributions due to its interactions with the density fields.
The latter is interpolated from the density values at the 
mesh points~\cite{Occam}.
Finally, we point out that, as observed in previous works
on similar systems~\cite{Denicola:16,Munao:18a}, in principle 
the incompressibility condition could cause the unphysical presence of PS chains
inside the NP. In order to avoid this effect, the NP density is modelled by
an analytical effective field, following a prescription reported in 
Ref.~\cite{Sides:06}. Specifically, the NP density field is represented
by a combination of two spline functions, one describing the density field
inside the NP core and the other 
preventing the external chains from overlapping
the NP. These functions are calculated by fixing four parameters: 
the NP radius $r_0$, the core
density $\phi_{core}$, the surface density $\phi_{max}$ and  the interval
$\delta r$ giving the width of the NP density profile ({\it i.e.} how fast the NP
density field goes from $\phi_{max}$ at $r = r_0$ to zero at
$r = r_0 + \delta r$). In our model we set the same parameters reported in 
Ref.~\cite{Munao:18a}, namely $r_0=2$, $\phi_{core}=100$,
$\phi_{max}=2$ and $\delta r=0.50$. The spline coefficients are then
obtained by imposing the continuity of the NP density and its derivative.

\subsection{Two- and three-body potentials of mean force}
For the calculation of the two-body PMF between two NPs,
for each system we have prepared
a set of 30 independent initial
configurations,
each one corresponding to a pair of NPs
placed at a fixed distance $d$ from each
other and embedded in the polymer matrix. 
A single simulation run has been performed from 
each starting configuration, 
where the NPs were allowed to freely rotate but
not to translate, in order to keep their distance fixed.
Forces $\bar F(d)$ on the centers of mass of the two NPs
have been computed at every 0.4 ps and
the convergence has been ensured by verifying that the average values of the
forces do not change anymore up to the first significant digit.
During the calculation of the two-body PMF, after a first simulation run of 
at least 60 ns, forces have been averaged over 
a subsequent production stage covering  the same amount of time. 
If not explicitly reported in the figures, error bars
corresponding to standard deviations are smaller than symbol sizes
of the corresponding curves.
The resulting PMF has been calculated according to the equation:
\begin{equation}\label{eq:PMF}
W(d)=-\int_{d}^{\infty} \bar{F}(r) dr \,,
\end{equation}
where $r$ is the interparticle distance, ranging in the $[d_{\rm min},
d_{\rm max}]$
interval. In our simulations, $d_{\rm min}=4$ nm and $d_{\rm max}=10$ nm;
therefore $d_{\rm min}$ is coincident with the NP diameter, while $d_{\rm max}$
indicates an interparticle distance where the potential can be confidently
assumed equal to zero. This $d_{max}$ 
is also used as the practical infinity in 
Eq.~\ref{eq:PMF}. In all simulations, distances are sampled with a step
of 0.2 nm and the numerical integration is performed by employing the trapezoidal rule.

\begin{figure*}[t!]
\begin{center}
\begin{tabular}{ccc}
\hspace{-1.0cm}
\includegraphics[width=5.9cm,angle=0]{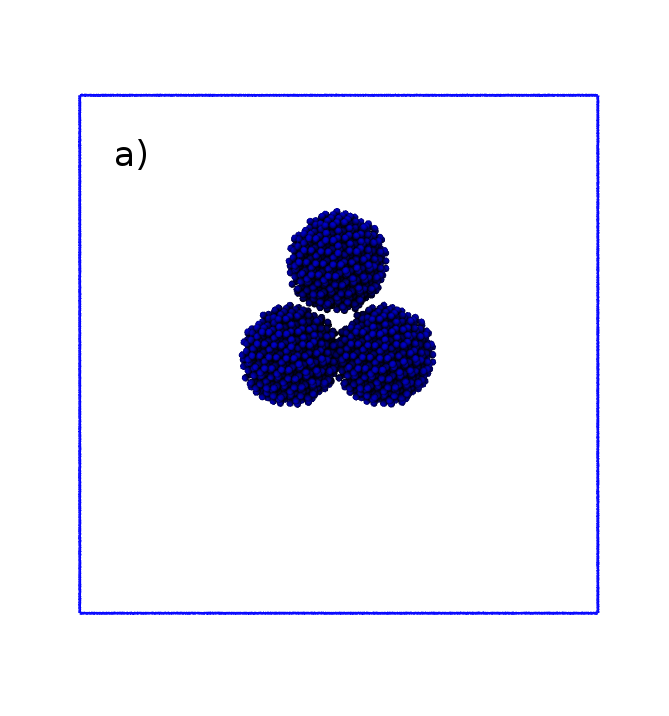}
\includegraphics[width=6.2cm,angle=0]{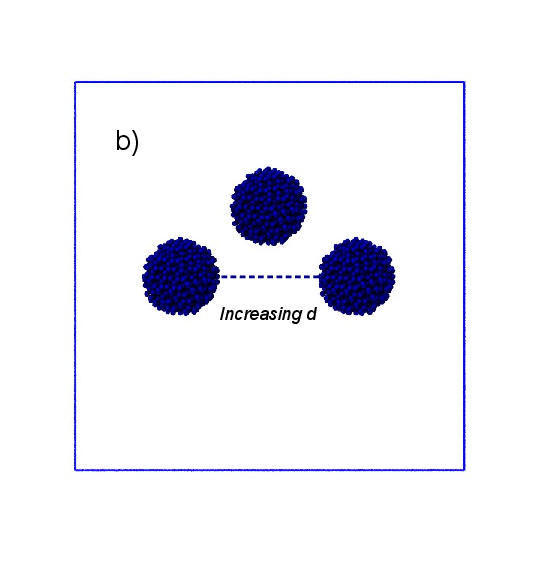}
\includegraphics[width=6.0cm,angle=0]{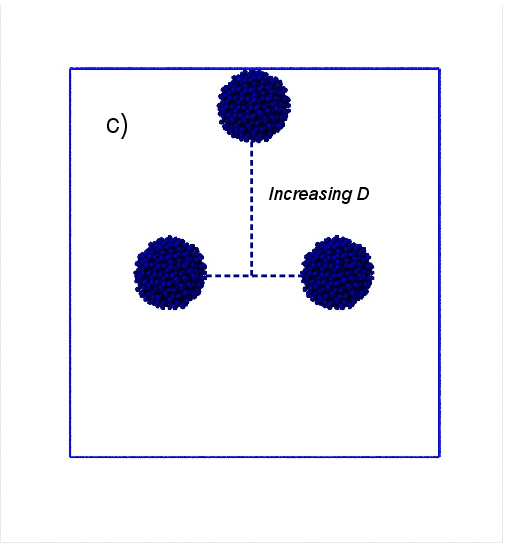}
\end{tabular}
\caption{Schematic procedure for the calculation of 
the three-body 
potential of mean force: starting from
a configuration where all nanoparticles are in close contact (a), the 
distance $d$ 
between the first two nanoparticles increases, 
leaving unchanged the position of the
third nanoparticle (b). The short or long-range nature of the three-body 
potential of mean force can be 
determined by increasing the distance $D$ between the third nanoparticle 
and the
center of the line joining the centers of mass of the first two 
nanoparticles (c).
}
\label{fig:F0a}
\end{center}
\end{figure*}

The procedure for calculating the three-body PMF requires a slightly
different approach: in this case the system is prepared including three NPs
at fixed distance from each other and surrounded by polymer chains. 
In analogy with the
two-body case, we increase the distance $d$ between the first two NPs, 
leaving the position of the third NP unchanged, and then we compute the forces
experienced by them. The three-body contribution
to the resulting PMF is maximum if the third NP is close to the others. 
In principle, 
also the distance $D$ between the third NP and the midpoint
of the line 
joining the first two NPs can be increased:  
this allows us
to estimate the short-range or long-range nature of the three-body 
interaction. Since we are interested in the computation of 
this interaction when the
third NP is close to the first two (when these effects are stronger) in our
simulations we have kept $D$ fixed to 4 nm.
The whole procedure is
schematically represented in Fig.~\ref{fig:F0a}.
In the calculation of the three-body
PMF, the convergence has been achieved after typically 
120 ns and the forces have
been averaged over a production run of 60 ns.  

%%%%%%%%%%%%%%%%%%%%%%%%%%%%%%%%%%%%%%%%%%%%%%%%%%%%%%
%%%%%%%%%%%%%%%%%%%%%%%%%%%%%%%%%%%%%%%%%%%%%%%%%%%%%
%%%%%%%%%%%%%%%%%%%%%%%%%%%%%%%%%%%%%%%%%%%%%%%%%%%%%%
\begin{table*}[t!]
\begin{center}
\caption{
Nanocomposite systems constituted by a single nanoparticle 
in a polystyrene matrix.
The grafting density $\rho_g$ is in unit of 
chains/nm$^2$. The box lengths are
$L_x=L_y=L_z=15.74$ nm.
The chain length is given in number of beads.
$BDI_f$ and $BDI_g$ are the bidispersity indices of free and grafted chains,
respectively.
}\label{tab:1NP}
\begin{tabular*}{1.05\textwidth}{@{\extracolsep{\fill}}ccccccccccccc}
\hline
\hline
& $\rho_g$ &  No$^{\circ}$ of & No$^{\circ}$ of 
& Short   & Long  & No$^{\circ}$ of & No$^{\circ}$ of &
 Short & Long & $BDI_g$ & $BDI_f$ \\
&  &  short & long  
&  grafted & grafted & short &  long &
 free & free &  &  \\
&  &  grafted & grafted
&  chains & chains & free &  free &
 chains & chains &  &  \\
&  &  chains & chains
&  length & length & chains &  chains &
 length & length &  &  \\
\hline
& 0  &  0 &  0 &  0 &  0 & 1070 & 0 & 20 & 0 & 0 & 1 \\
& 0  &  0 &  0 &  0 &  0 & 840 & 23 & 20 & 200 & 0 & 2.36  \\
& 0  &  0 &  0 &  0 &  0 & 100 & 19 & 20 & 1000 & 0 & 5.14  \\
& 0.1  & 0 & 5 &  0 & 80 & 1070 & 0 & 20 & 0 & 1 & 1 \\
& 0.1  & 0 & 5 &  0 & 80 & 840 & 23 & 20 & 200 & 1 & 2.36 \\
& 0.1  & 0 & 5 &  0 & 80 & 100 & 19 & 20 & 1000 & 1 & 5.14 \\
& 0.1  & 1 & 4 &  10 & 80 & 1005 & 0 & 20 & 0 & 2.36 & 1 \\
& 0.1  & 1 & 4 &  5 & 80 & 1007 & 0 & 20 & 0 & 3.25 & 1 \\
& 0.5  & 0 & 25 &  0 & 80 & 877 & 0 & 20 & 0 & 1 & 1 \\
& 0.5  & 0 & 25 &  0 & 80 & 177 & 69 & 20 & 200 & 1 & 2.36 \\
& 0.5  & 0 & 25 &  0 & 80 & 84 & 16 & 20 & 1000 & 1 & 5.14 \\
& 0.5  & 5 & 20 &  10 & 80 & 991 & 0 & 20 & 0 & 2.36 & 1 \\
& 0.5  & 5 & 20 &  5 & 80 & 1001 & 0 & 20 & 0 & 3.25 & 1 \\
\hline
\end{tabular*}
\end{center}
\end{table*}
%%%%%%%%%%%%%%%%%%%%%%%%%%%%%%%%%%%%%%%%%%%%%%%%%%%%%%

%%%%%%%%%%%%%%%%%%%%%%%%%%%%%%%%%%%%%%%%%%%%%%%%%%%%%%
\subsection{System preparation}

Since the effective interactions between NPs are mediated by 
the interactions with polymer chains, the calculation of the PMF has been
systematically supported by simulations 
of systems comprising 
a single NP (ungrafted or grafted) in a PS matrix, which contains
polymer chains of different lengths. 
In a previous work~\cite{Munao:18a} we have investigated the
role played by the molecular weight of monodisperse free and grafted chains on 
the NP-NP interactions. In the present work  
we focus on the role played by bidispersity
and grafting density. 
For such an aim we have
introduced a bidispersity index ($BDI$) defined as:
\begin{equation}
BDI=M_w/M_n
\end{equation}
where $M_w$ and $M_n$ are the weight average molecular weight and the number
average molecular weight, respectively, and are defined as:
\begin{equation}\label{eq:Mn}
M_w=\frac{\sum_i N_i M_i^2}{\sum_i N_i M_i}; \qquad 
M_n=\frac{\sum_i N_i M_i}{\sum_i N_i} \,,
\end{equation}
where 
$N_i$ and $M_i$
indicate the number of chains with a given molecular weight and the number of
beads belonging to each of these chains.
The systems are listed in Tab.~\ref{tab:1NP}, along with
the grafting densities, the number and molecular weight of free and grafted
chains and  
two different bidispersity
indices $BDI_f$ and $BDI_g$, which take into account the
bidispersity of free and grafted chains, respectively.

%%%%%%%%%%%%%%%%%%%%%%%%%%%%%%%%%%%%%%%%%%%%%%%%%%%%%%
\begin{table*}[t!]
\begin{center}
\caption{Nanocomposite systems investigated for the calculation of the
two-body potential of mean force.
The grafting density $\rho_g$ is in unit of
chains/nm$^2$. The box lengths 
are $L_x=22$ nm, $L_y=L_z=12.5$ nm.
The chains length is given in number of beads.
}\label{tab:2PMF}
\begin{tabular*}{1.05\textwidth}{@{\extracolsep{\fill}}ccccccccccccc}
\hline
\hline
& $\rho_g$ &  No$^{\circ}$ of & No$^{\circ}$ of 
& Short   & Long  & No$^{\circ}$ of & No$^{\circ}$ of &
 Short & Long & $BDI_g$ & $BDI_f$ \\
&  &  short & long  
&  grafted & grafted & short &  long &
 free & free &  &  \\
&  &  grafted & grafted
&  chains & chains & free &  free &
 chains & chains &  &  \\
&  &  chains & chains
&  length & length & chains &  chains &
 length & length &  &  \\
\hline
& 0  &  0 &  0 &  0 &  0 & 1044 & 0 & 20 & 0 & 0 & 1 \\
& 0  &  0 &  0 &  0 &  0 & 300 & 75 & 20 & 200 & 0 & 2.36  \\
& 0  &  0 &  0 &  0 &  0 & 950 & 2 & 20 & 1000 & 0 & 5.14  \\
& 0.1   &  0 &  5 & 0 & 80 &  995 & 0&  20 & 0 & 1 & 1 \\
& 0.1   &  0 &  5 & 0 & 80 &  840 & 23&  20 & 200 & 1 & 2.36 \\
& 0.1   &  0 &  5 & 0 & 80 &  100 & 19&  20 & 1000 & 1 & 5.14 \\
& 0.1   &  3 &  2 & 13 & 180 &  995 & 0&  20 & 0 & 2.00 & 1 \\
& 0.1   &  4 &  1 & 10 & 80 &  1005 & 0&  20 & 0 & 2.36 & 1 \\
& 0.1   &  4 &  1 & 5 & 80 &  1007 & 0&  20 & 0 & 3.25 & 1 \\
& 0.1   &  4 &  1 & 80 & 1000 &  903 & 0&  20 & 0 & 2.94 & 1 \\
& 0.5   &  0 & 25 & 0 & 80 & 854 & 0 & 20 & 0 & 1 & 1 \\
& 0.5   &  0 & 25 & 0 & 80 & 187 & 69 & 20 & 200 & 1 & 2.36 \\
& 0.5   &  0 & 25 & 0 & 80 & 84 & 16 & 20 & 1000 & 1 & 5.14 \\
& 0.5   &  20 & 5 & 40 & 240 & 854 & 0 & 20 & 0 & 2.00 & 1 \\
& 0.5   &  20 & 5 & 10 & 80 & 991 & 0 & 20 & 0 & 2.36 & 1 \\
& 0.5   &  20 & 5 & 5 & 80 & 1001 & 0 & 20 & 0 & 3.25 & 1 \\
\hline
\end{tabular*}
\end{center}
\end{table*}
%%%%%%%%%%%%%%%%%%%%%%%%%%%%%%%%%%%%%%%%%%%%%%%%%%%%%%
The total collection of investigated systems for the calculation of 
the two-body PMF 
is reported in Tab.~\ref{tab:2PMF}: note that in this case the simulation
box 
is not cubic but is elongated along the direction joining
the centers of the two NPs.
In all bidisperse systems the polymer density is 
equal (or very close) to that of monodisperse cases. However, since the 
chains (free or grafted) follow a bimodal distribution, the number 
average molecular
weight $M_n$ (see Eq.~\ref{eq:Mn}) is not the same 
as in the unimodal case. 
For bidisperse
free chains this circumstance is not expected to influence the two-body PMF, 
in agreement with 
recent simulation~\cite{Munao:18a} and theoretical~\cite{Hooper:04} studies 
where it has been demonstrated that
the two-body PMF is weakly dependent on  
the molecular weight of the 
free chains $M_n$. For bidisperse grafted chains we have first 
kept the $M_n$ constant, in order to limit  the effects
only due to the bidispersity. Then, also $M_n$  has been
varied, allowing us to explore a large variety of behaviors. 
We have verified that the radius of the NP (corresponding to 2 nm)
is equivalent to the gyration radius of chains containing 80 
beads~\cite{Munao:18a}. Hence, the mean size of the longest chains simulated 
(containing 1000 beads) is considerably larger than the size 
of a single NP. \\ 
The total interaction between the 
pair of NPs has 
two contributions: the first is the direct interaction among
the NP cores, which is 
modeled by a Hamaker potential~\cite{Hamaker:37}, parameterized 
on the basis of
atomistic simulations in Ref.~\cite{Munao:18}. Specifically, in that work the
NP-NP interactions were calculated for ungrafted silica NPs of different sizes;
in all cases the Hamaker potential was found to agree very well with the
atomistic simulations, indicating that it accurately reproduces the NP
core-core attractive interaction.
The second contribution to the total interaction is 
given by the surrounding polymer chains, calculated 
by performing molecular
simulations according to the procedure described in Section 
{\bf Two- and three-body potentials of mean force}. 
More details on this approach, successfully applied to 
silica-polystyrene nanocomposites with monodisperse PS chains, can be
found in Refs.~\cite{Munao:18,Munao:18a}.
In order to discriminate between the effects due to the bidispersity 
of free and grafted chains, as well as on the different behaviors between
ungrafted and grafted NPs, we have separately discussed these cases in
different subsections.

\section{Results and discussion}
\subsection {Increasing the bidispersity of free chains: the ungrafted case}

%%%%%%%%%%%%%%%%%%%%%%%%%%%%%%%%%%%%%%%%%%%%%%%%%%%%%%
\begin{figure}[t!]
\begin{center}
\includegraphics[width=8.0cm,angle=0]{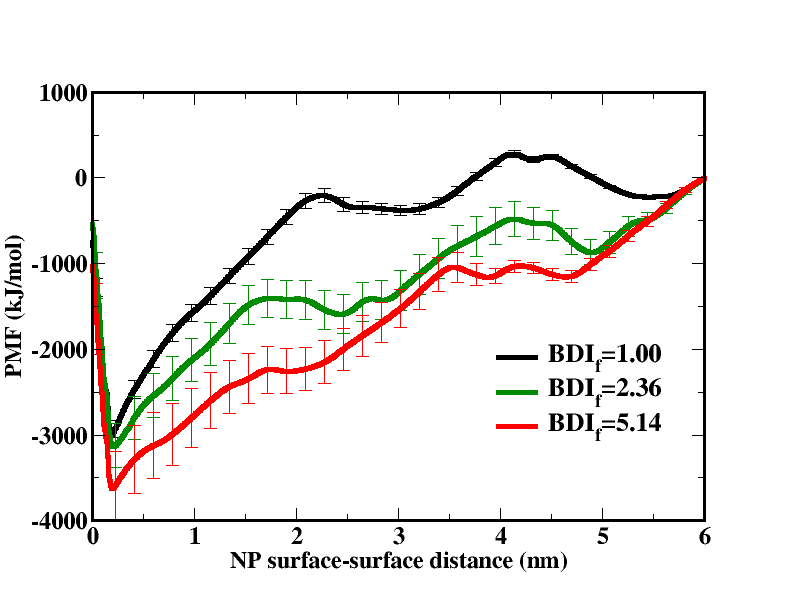}
\caption{Effect of increasing the bidispersity index of the free chains 
on the potential of mean force between two ungrafted nanoparticles.}
\label{fig:F1}
\end{center}
\end{figure}
%%%%%%%%%%%%%%%%%%%%%%%%%%%%%%%%%%%%%%%%%%%%%%%%%%%%%

The PMF between a pair of ungrafted NPs for different $BDI_f$ is reported
in Fig.~\ref{fig:F1}: for the monodisperse case (black line), 
the effective interaction
is remarkably attractive, as already observed in Ref.~\cite{Munao:18a}.
The attraction, in particular at shorter NP-NP distance, 
is due to the chain 
confinement, since
only a small number of chains with elongated conformations fit between
the two NPs. Conversely, for larger interparticle separations, a progressively
higher number of free chains can be arranged therein until this number
becomes comparable with the bulk and the attraction vanishes.
The increase of the bidispersity
of free chains, obtained by mixing short (20 beads) and long
(200 and 1000 beads, see Tab.~\ref{tab:2PMF}) chains 
leads to a slight 
increase of the attraction between the NPs. 
In addition, we note that the oscillating behavior, observed
for $BDI_f=1$ is progressively suppressed.
The physical length which governs the wavelength of the 
oscillations is the NP radius~\cite{Munao:18a}.
As explained in previous theoretical~\cite{Hooper:04,Hooper:05,Yeth:11} and
simulation~\cite{Bedrov:03,Munao:18a} studies,
oscillations develop for moderate NP-NP attractions, where the 
regions of perturbed polymer density around each
NP can overlap. 
The progressive disappearance of the oscillations 
indicates the breakdown of this mechanism, and the simultaneous 
onset of a stronger attraction between the NPs.
All these effects can be entirely ascribed to the increase of
$BDI_f$ since, as previously anticipated, the free chains $M_n$
does not influence the two-body PMF~\cite{Hooper:04,Munao:18a}.

\begin{figure}[t!]
\begin{center}
\includegraphics[width=8.0cm,angle=0]{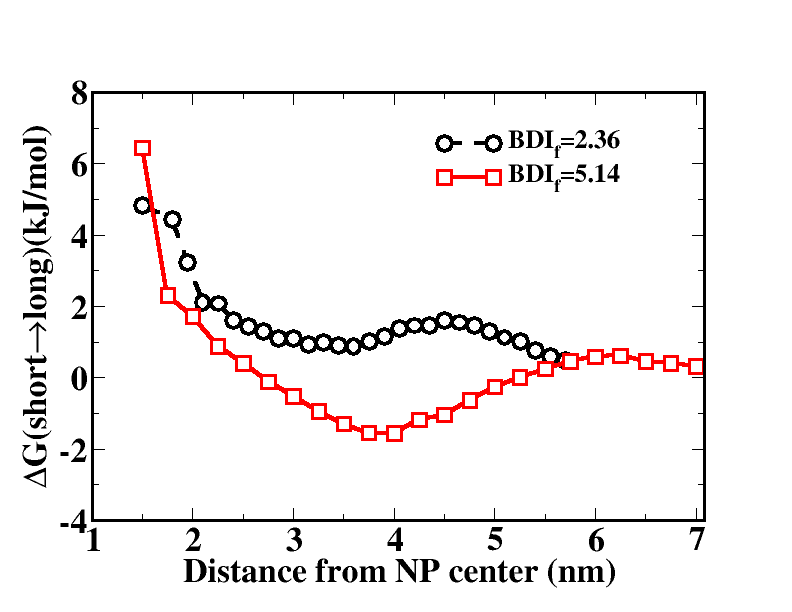}
\caption{Gibbs free energy for the exchange of a monomer belonging to a short 
free chain with a monomer belonging to a long free chain as a function of the 
distance between
the chains and the center of a ungrafted nanoparticle. Two different 
bidispersity indices of the free chains (reported in the legend) 
are considered.} 
\label{fig:Fdelta}
\end{center}
\end{figure}

\begin{figure*}[t!]1
\begin{center}
\begin{tabular}{cc}
\hspace{-2.0cm}
\includegraphics[width=9.0cm,angle=0]{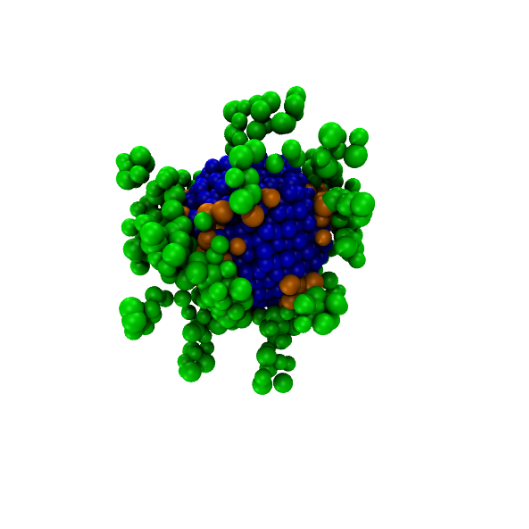}
\hspace{-2.0cm}
\includegraphics[width=9.0cm,angle=90]{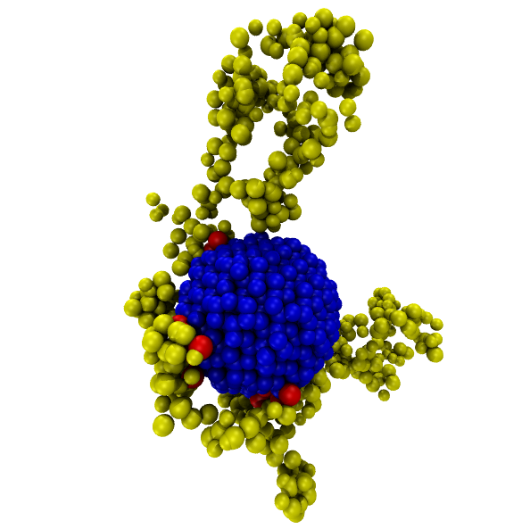}
\end{tabular}
\vspace{-2.0cm}
\caption{Local arrangements of short (left) and long (right) free chains around
a ungrafted NP for a bidispersity index of the free chains set to 2.36. 
Only chains with at least one bead placed 
within 0.5 nm from the nanoparticle surface are reported. 
Short and long chains are 
depicted in green and yellow, while the beads close to the nanoparticle
surface are given in orange and red, respectively. }
\label{fig:Fsnap}
\end{center}
\end{figure*}

In order to provide an explanation of the increase of the attraction with
the bidispersity, we have investigated the arrangement of short and long
chains around a single NP by calculating their monomer 
number density
for different values of $BDI_f$.
%
%%%%%%%%%%%%%%%%%%%%%%%%%%%%%%%%%%%%%%%%%%%%%%%%%%%%%%
\begin{figure}[t!]
\begin{center}
\includegraphics[width=9.0cm,angle=0]{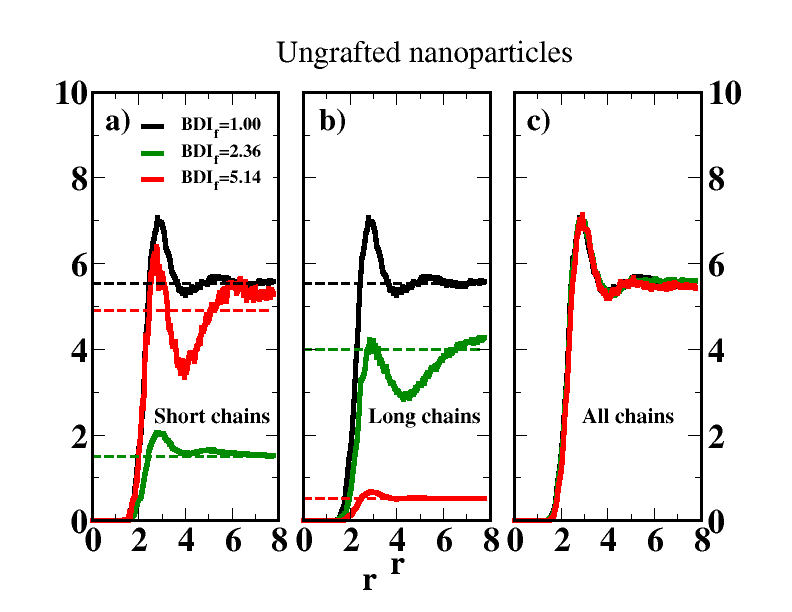}
\caption{Monomer number density of
polymer chains as a function of their distance from the center of 
an 
ungrafted nanoparticle upon
increasing the bidispersity index of the free chains.
Short (a), long (b) and all (c) chains are separately reported.
Broken horizontal lines indicate the density levels that 
would correspond to the short and long chains if the mixture was homogeneous.
}
\label{fig:F1a}
\end{center}
\end{figure}
%%%%%%%%%%%%%%%%%%%%%%%%%%%%%%%%%%%%%%%%%%%%%%%%%%%%%
%
Results are reported in Fig.~\ref{fig:F1a}: 
in particular, we separately show the cases corresponding to short 
(panel (a)), 
long (panel (b)) and all (panel (c)) chains. 
It is worth pointing out that long chains
comprise 200 beads if $BDI_f=2.36$ and 1000 beads if $BDI_f=5.14$ 
(see Tab.~\ref{tab:1NP}).
The monomer number density for the monodisperse case exhibits 
the maximum value 
for a distance from the NP center of $\approx$ 3 nm, followed by  
the minimum value and the second maximum 
for a distance of $\approx$ 5 nm, finally
converging to its bulk value.
The first peak in the monomer number density of short chains is still observed 
for $BDI_f$=2.36 and the density profile becomes even more structured for
$BDI_f$=5.14. This trend is reversed
for the long chains (panel (b)), with the 
density profile becoming progressively less structured upon increasing
$BDI_f$.
The exchange in the order of short and long chains is
undoubtedly a consequence of the different 
numbers of monomers belonging to the
two kinds of chains, but nevertheless the emerging picture (suggesting 
a shell structure, with the short chains arranged close to the NP surface 
and the long chains around them) seems genuine.
The broken horizontal lines shown in panels (a) and (b) indicate
the density values corresponding to homogeneous distributions of short and 
long chains. The monomer number densities of both short and long chains
do not attain their homogeneous values 
even for distances from the NP center as large as 8 nm.
In comparison with a multiscale simulation study of
PS-SiO$_2$ nanocomposites~\cite{Mathioudakis:16}, where a layering of 
monodisperse chains was observed, in our study the chain layering extends 
until higher distances from the NP surface. This effect can be ascribed
to the bidispersity in the chain length.
In addition,
if we consider the collective contribution given by short and long chains
together (panel (c)) we find no effects due to $BDI_f$: this suggests that
the structuring of the number density 
for short chains is compensated by the decrease
of that for long chains and vice versa.
In order to quantify
the difference of the chain arrangements around the NP surface, 
we have calculated the Gibbs free energy
$\Delta G(r)$ for the replacement of a monomer of a 
short chain 
with that of a long chain in the
coordination environment of a short chains, following the prescription
reported in Ref.~\cite{Milano:04}:
\begin{equation}
{\rm ln}\frac{g_{{\rm long}}(r)}{g_{{\rm short}}(r)}=\frac{-{\it {\Delta G}(r)}({\rm short} \rightarrow {\rm long})}{RT} \,,
\end{equation}
where $g_{{\rm short}}(r)$ and $g_{{\rm long}}(r)$ are 
the two radial distribution functions for monomers belonging to 
short and long chains around the NP, respectively, and $r$ is the distance
from the NP center. 
In Fig.~\ref{fig:Fdelta} we report the values of 
$\Delta G(r)({\rm short \rightarrow long})$ 
for $BDI_f=2.36$ and 5.14: 
in both cases $\Delta G(r)$ is strictly positive in proximity of the NP
surface. Upon increasing the distance, for $BDI_f=2.36$ $\Delta G(r)$ 
remains positive, whereas for $BDI_f=5.14$ the 
minimum is found when the
distance from the NP center is 4 nm. 
We also note that, due to the softness of the NP-PS interaction,
$\Delta G(r)$ shows non-zero 
values also for distances from the NP center
less than the NP radius. A similar behavior was also observed in a simulation 
study of a CG model for athermal all-polystyrene 
nanocomposite~\cite{Vogiatzis:11a}: in particular, the authors observed that
when the radius of gyration of chains is comparable in size to the
nanoparticle the centers of mass of the chains 
may penetrate the interior of the NP,
as chains engulf the NP. 
According to the picture provided 
by Figs.~\ref{fig:F1a}-\ref{fig:Fdelta}, 
it emerges that, when the PS chains 
are bidisperse, replacing a monomer belonging to short chain with one belonging
to a long chain carries a Gibbs free energy penalty; hence there is a 
tendency of short chains to be arranged close to
the NP surface, with the long chains just outside this first shell. 
%%%%%%%%%%%%%%%%%%%%%%%%%%%%%%%%%%%%%%%%%%%%%%%%%%%%%%
%%%%%%%%%%%%%%%%%%%%%%%%%%%%%%%%%%%%%%%%%%%%%%%%%%%%%
If two
NPs come in close contact (see Fig.~\ref{fig:F1})
this configuration can no longer be realized, since
the long chains can not be arranged in the narrow space between the NPs. As a
consequence, each NP has to ``desorb'' short chains from the first shell,
in order to put them at the interface with the other NP. This desorption
causes the increase of the attraction between the NPs upon increasing $BDI_f$.
This picture can be qualitatively confirmed by comparing the 
number of monomers belonging to short and long chains around a single 
``unperturbed'' NP with that found when two NPs come in close contact.
The difference in the composition of the first shell (from 0 to 0.5 nm
from the NP surface) around the NP of monomers belonging to short and long
chains can be estimated; for $BDI_f=2.36$ the number of monomers 
belonging to short and long chains changes from 74 and 12 for a single NP
to 18 and 53 for a pair of NPs. Similarly, this number changes from 67 and 10
to 9 and 70 for $BDI_f=5.14$. Hence, there are 103 substitutions in the first
case and 118 in the second case; by multiplying these values for $\Delta G(r)$
we obtain 463 kJ/mol for $BDI_f=2.36$ and 767 kJ/mol for $BDI_f=5.14$, 
which qualitatively follows the increase
of the attraction between the two NPs.
%%%%%%%%%%%%%%%%%%%%%%%%%%%%%%%%%%%%%%%%%%%%%%%%%%%%%
Interestingly, a similar behavior is described in Ref.~\cite{Martin:14}, where
the authors investigate the effect of the increase of 
$BDI_f$ on the effective 
interactions between grafted NPs. In their case the molecular weight 
of the grafted
chains is systematically lower than that of free chains, regardless of their
bidispersity; as a consequence, the resulting interaction is attractive
since in all cases the length of the free chains is higher than 
the length of the grafted chains~\cite{Green:11,Martin:13a,Kumar:13}.
The authors observe that short chains preferentially wet the grafted layer
in comparison with the long chains, 
showing an higher correlation with the NPs at short distances.
This can be ascribed to an higher gain in the 
entropy of mixing between grafted and
short chains than that between grafted and long chains. Indeed, it is known
that for monodisperse systems the wetting of the grafted layer decreases 
if the length of the free chains 
increases~\cite{Green:11,Martin:13,Trombly:10}; 
hence, since the length of the free chains is higher than 
the length of the grafted chains,
short chains preferentially wet the 
grafted layer more than long chains. 
In our case the NP is ungrafted and the tendency of short
chains to wet the NP surface more than the long chains appears even stronger.
The different arrangements of short and long chains around a single ungrafted 
NP for $BDI_f=2.36$ are reported in Fig.~\ref{fig:Fsnap}: 
more specifically, we show only chains which at least one bead placed within a
distance of 0.5 nm from the NP surface. It emerges that there are 
typically 30
short chains and 3 long chains with 119 and 16 beads close to the NP surface,
respectively. Hence, each long chain, on average, has 5.33 beads 
which fall
within 0.5 nm from the NP surface; for short chains this number is 3.96. 
This means that the long chains are multiply bent when approaching the
NP surface and such configurations are entropically unfavoured in comparison
to the short chains that do not experience similar arrangements.
%%%%%%%%%%%%%%%%%%%%%%%%%%%%%%%%%%%%%%%%%%%%%%%%%%%%%%
%%%%%%%%%%%%%%%%%%%%%%%%%%%%%%%%%%%%%%%%%%%%%%%%%%%%%%

We have further characterized the NP-PS interface by 
calculating the work of adhesion $W_{ad}$ between the PS melt and 
a single silica NP. 
In our approach we calculate $W_{ad}$ as:
\begin{equation}
W_{ad}=\frac{\sum_N V_{NP-PS}(\bf r)}{A_{NP}} \,,
\end{equation}
where $A_{NP}$ is the surface area of the NP and 
\begin{equation}
V_{NP-PS}({\bf r})=k_B T \chi_{KK'} \Phi_{K'}({\bf r}) \,,
\end{equation}
which corresponds to the first term of the right hand side of Eq.~\ref{eq:pot}.
Taking into account the roughness of the NP surface, we have calculated its 
area by means of the GROMACS~\cite{Gromacs} tool for the solvent access 
surface area, with a value of $r_{probe}$ corresponding to the radius of a 
polymer bead ({\it i.e.} 0.35 nm). By following this prescription we have 
obtained 90 nm$^2$ for the NP area.
According to this procedure the work of adhesion is $W_{ad}=58.89$ mJ/m$^2$ for
$BDI_f$=1.00, $W_{ad}=60.55$ mJ/m$^2$ for $BDI_f=2.36$ and 
$W_{ad}=67.78$ mJ/m$^2$
for $BDI_f=5.14$. Hence we document a progressively stronger adhesion of 
PS chains onto the NP surface upon increasing $BDI_f$. 
Such estimations satisfactorily agree
with the experimental value~\cite{Mortezaei:11} 
of $W_{ad}=70$ mJ/m$^2$ (found for monodisperse systems), especially considering
that the proposed CG model has been developed for reproducing the density
profiles rather than interfacial properties of silica-PS composites.

\subsection {Increasing the bidispersity of free chains: the grafted case}

\begin{figure*}[t!]
\begin{center}
\begin{tabular}{cc}
\includegraphics[width=8.0cm,angle=0]{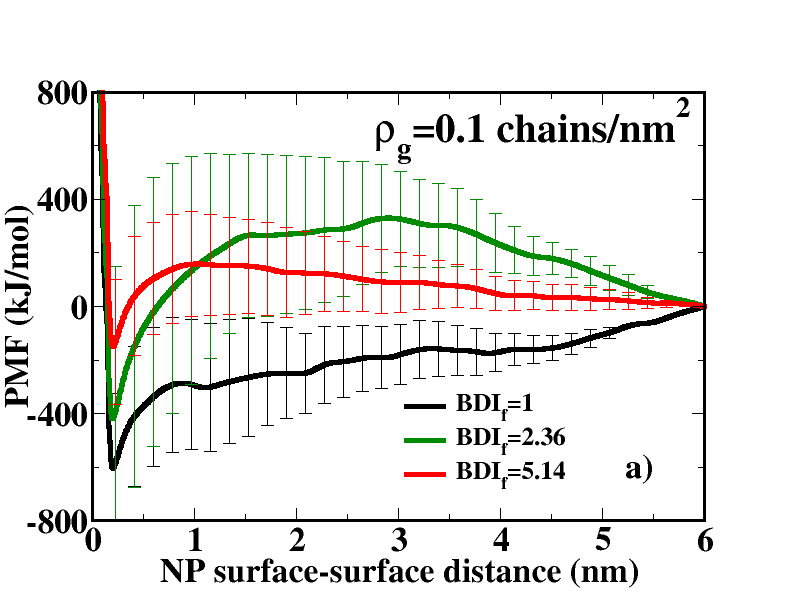}
\includegraphics[width=8.0cm,angle=0]{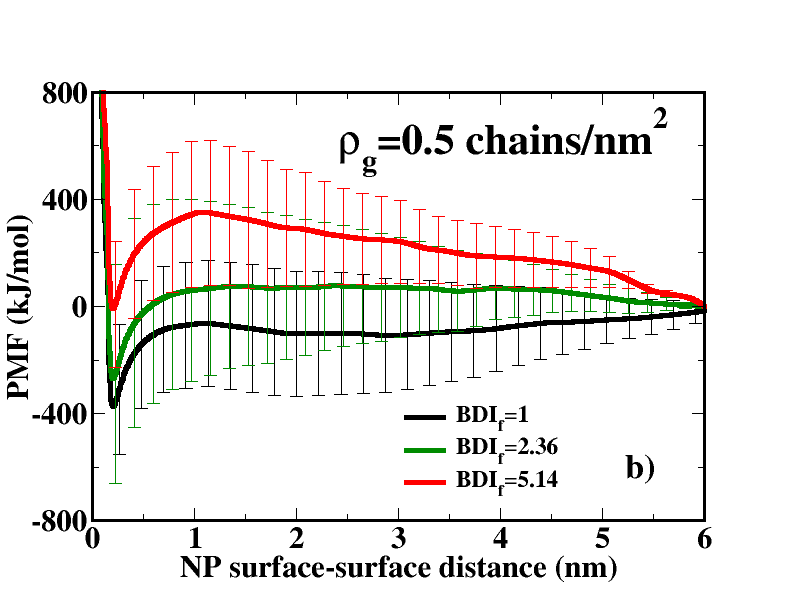}
\end{tabular}
\caption{Effect of increasing the bidispersity index of free chains 
on the potential of mean
force between two grafted nanoparticles 
with grafting density $\rho=0.1$ (a) and 0.5 (b) chains/nm$^2$. 
In all cases the grafted chains are monodisperse.}
\label{fig:F2}
\end{center}
\end{figure*}

Upon grafting the NPs, the scenario significantly
changes, as observed by calculating the PMF between a pair of 
grafted NPs with a grafting density of 0.1 chains/nm$^2$ (see
Fig.~\ref{fig:F2}a). We first investigate the effects due to the increase of
$BDI_f$, leaving $BDI_g$ unchanged.
For the monodisperse case, the PMF is attractive
in the whole investigated range of interparticle distances, 
even if the strength of
the attraction is considerably lower than the ungrafted case. 
This is due to the grafted chains that, as described in our previous 
work~\cite{Munao:18a}, introduce repulsive contributions 
able to stabilize the
whole nanocomposite with respect to the aggregation. 
Indeed we have shown in the same work that the increase of  
either the grafting density or the length of the grafted chains improves 
the miscibility of the
NPs with the PS melt. The same behavior was found also in a simulation study
of silica-PS nanocomposites~\cite{Vogiatzis:13} performed by means of a 
Monte Carlo methodology based on polymer mean field theory. 
Upon increasing $BDI_f$ the repulsion between NPs increases in turn,
especially in the intermediate range of interparticle distances.
In addition, we note the existence of a crossover between the behaviors 
of the two bidisperse cases: for interparticle distances closer than 1 nm,
the repulsion monotonically increases with $BDI_f$, whereas 
for larger distances
the repulsion is higher for $BDI_f=2.36$ than for $BDI_f=5.14$. 
%
%%%%%%%%%%%%%%%%%%%%%%%%%%%%%%%%%%%%%%%%%%%%%%%%%%%%%%
%%%%%%%%%%%%%%%%%%%%%%%%%%%%%%%%%%%%%%%%%%%%%%%%%%%%%

Upon increasing the grafting density to $\rho_g=0.5$ chains/nm$^2$
(Fig.~\ref{fig:F2}b) the two-body PMF similarly behaves:
in the unimodal distribution the effective interaction
is attractive in the whole range of
interparticle separations and becomes 
progressively more repulsive upon increasing $BDI_f$.
However, in this case the repulsion monotonically increases with the
bidispersity, without showing any crossover.
Hence, it emerges that the increase of 
$BDI_f$ favours a good
dispersion of the NPs for both the grafting densities. 

%%%%%%%%%%%%%%%%%%%%%%%%%%%%%%%%%%%%%%%%%%%%%%%%%%%%%%
\begin{figure*}[t!]
\begin{center}
{\large {Increasing $BDI_f$}} \\
{\LARGE {$\downarrow$}} \\
\vspace{-0.3cm}
\includegraphics[width=8.0cm,angle=0]{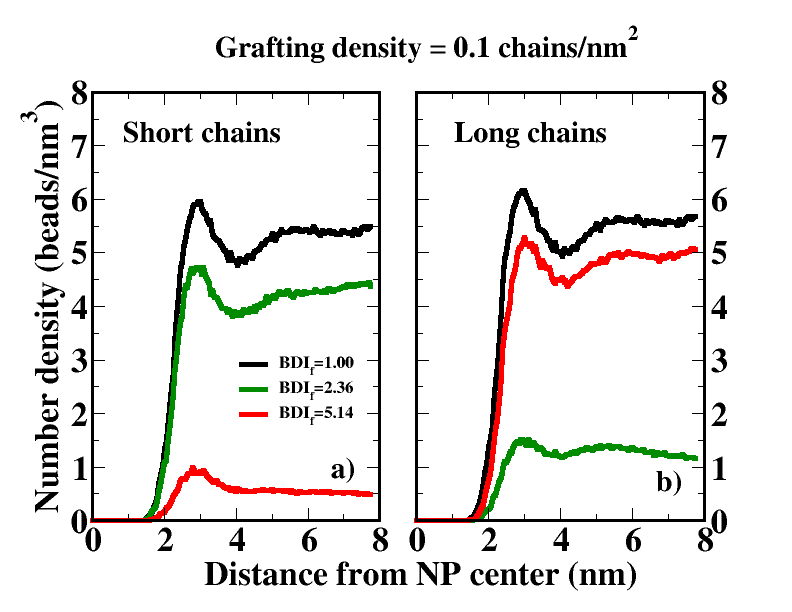}
\includegraphics[width=8.0cm,angle=0]{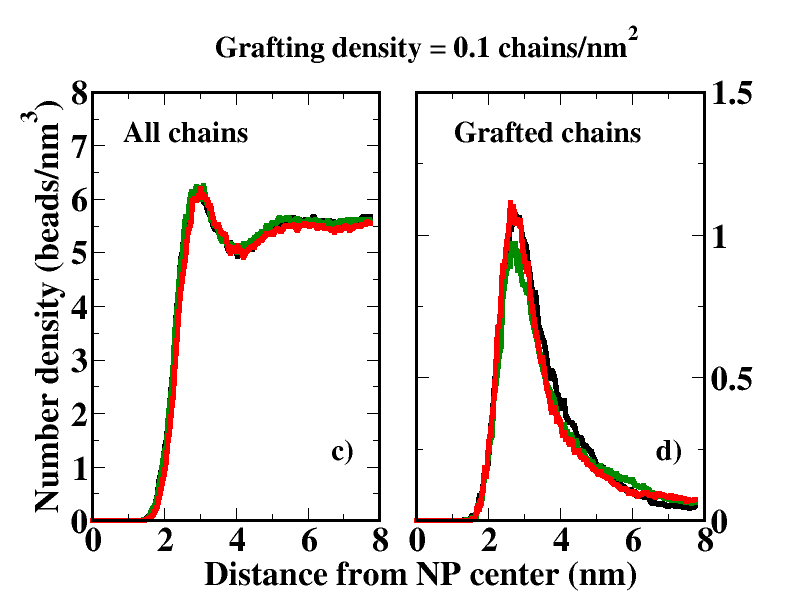} \\
\includegraphics[width=8.0cm,angle=0]{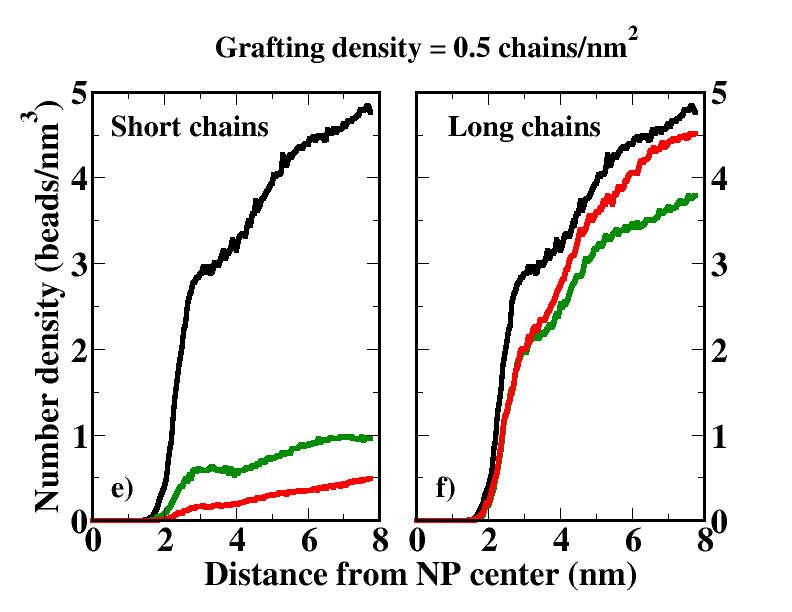} 
\includegraphics[width=8.0cm,angle=0]{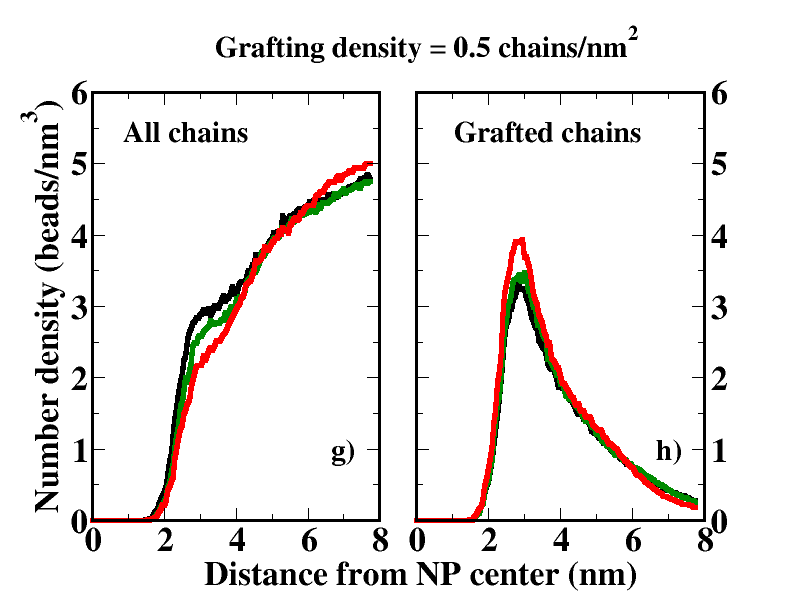} 
\caption{Radial number density of monomers
belonging to free and grafted chains.
Two different grafting densities, namely
$\rho_g=0.1$ (top panels) and 0.5 (bottom panels) chains/nm$^2$ are reported. 
Short, long and all free chains are separately shown.
}
\label{fig:F1b}
\end{center}
\end{figure*}
%%%%%%%%%%%%%%%%%%%%%%%%%%%%%%%%%%%%%%%%%%%%%%%%%%%%%
As previously done for ungrafted NPs, a comparison with the behavior of the
monomer number density
of free and grafted chains around a single NP can be helpful for understanding
the behavior of the effective interactions. In the top panels of 
Fig.~\ref{fig:F1b} we show the bead number density profiles for 
$\rho_g=0.1$ chains/nm$^2$: 
in comparison with the ungrafted case (see Fig.~\ref{fig:F1a}) 
the density profiles of both short and long chains oppositely
behave upon increasing $BDI_f$. As a consequence, the two-body PMF is
progressively more repulsive: 
this suggests that, beside the repulsion between
grafted chains already existing in the monodisperse case, 
there is an extra repulsion
between grafted chains and the short chains lying at the interface 
between the two NPs. In order to confirm this assumption, we have checked the
behavior of short and long chains by looking at different configurations 
in proximity of the NPs during the 
calculations of the two-body PMF; we have verified that for short
intermolecular distances, the long chains can hardly be arranged between the 
two NPs and short chains are progressively pushed away from the NP surfaces.
This behavior is not observed for ungrafted NPs and is due to the repulsive
interaction between free and grafted chains.
Similar effects have been 
observed also in simulation studies of atomistic models of silica
NPs in a PS~\cite{Ndoro:11,Ndoro:12} or PMMA~\cite{Eslami:13} melts.
It is also worth pointing out that, unlike the systems investigated 
in Ref.~\cite{Martin:14} and previously discussed, in this case the 
length of the free chains is not always higher than the 
length of the grafted chains:
this explains the different
PMF observed.
 
For $\rho_g=0.5$ chains/nm$^2$ (bottom panels of Fig.~\ref{fig:F1b}) the
profiles of both short and long chains significantly
change: because of the higher number of grafted chains, the space available
for free chains is substantially reduced, regardless of their molecular weight.
Hence, no peak is found in their distributions, which monotonically 
approach the bulk value. 
However, the trends of short and long chains upon increasing
$BDI_f$ are the same as that 
observed for $\rho_g=0.1$ chains/nm$^2$ and hence the
corresponding two-body PMF (reported in
Fig.~\ref{fig:F2}b) are progressively more repulsive for higher
bidispersities.

\subsection {Increasing the bidispersity of grafted chains}

%%%%%%%%%%%%%%%%%%%%%%%%%%%%%%%%%%%%%%%%%%%%%%%%%%%%%%
\begin{figure*}[t!]
\begin{center}
{\large {Increasing $BDI_g$}} \\
{\LARGE {$\downarrow$}} \\
\vspace{-0.3cm}
\includegraphics[width=8.0cm,angle=0]{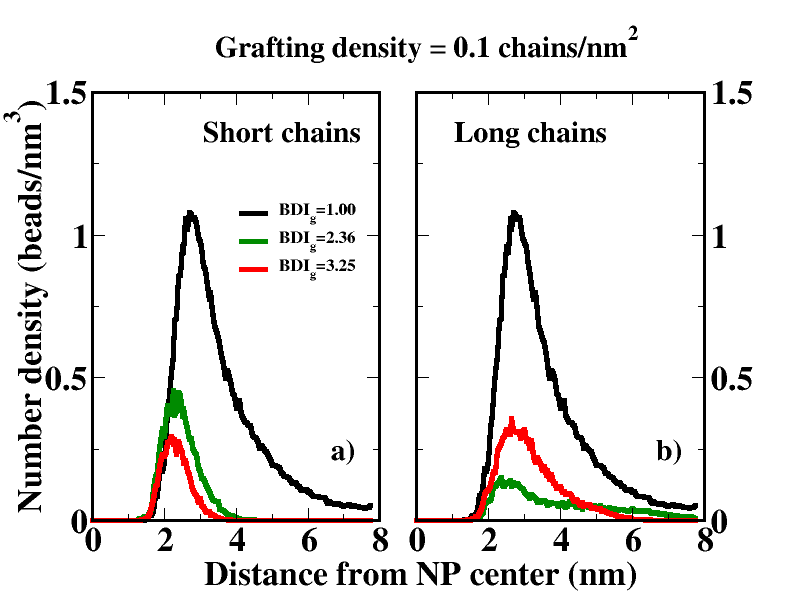}
\includegraphics[width=8.0cm,angle=0]{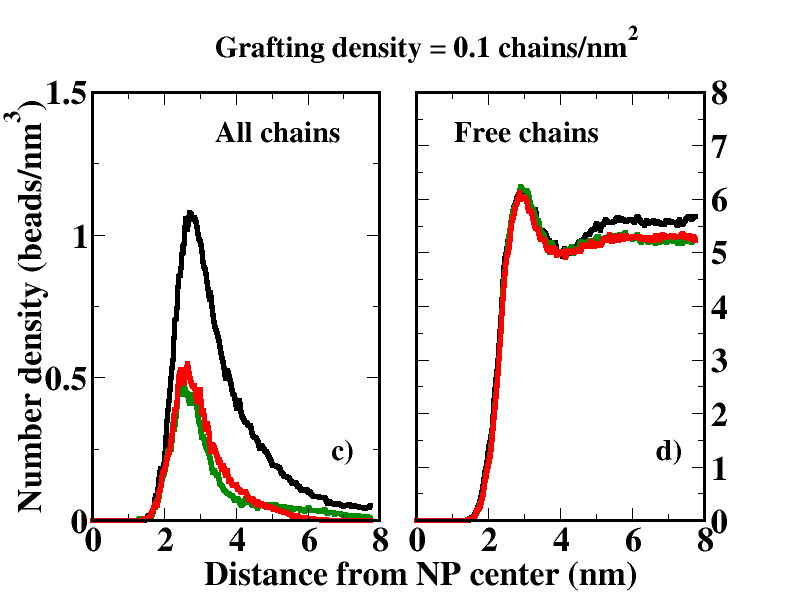} \\
\includegraphics[width=8.0cm,angle=0]{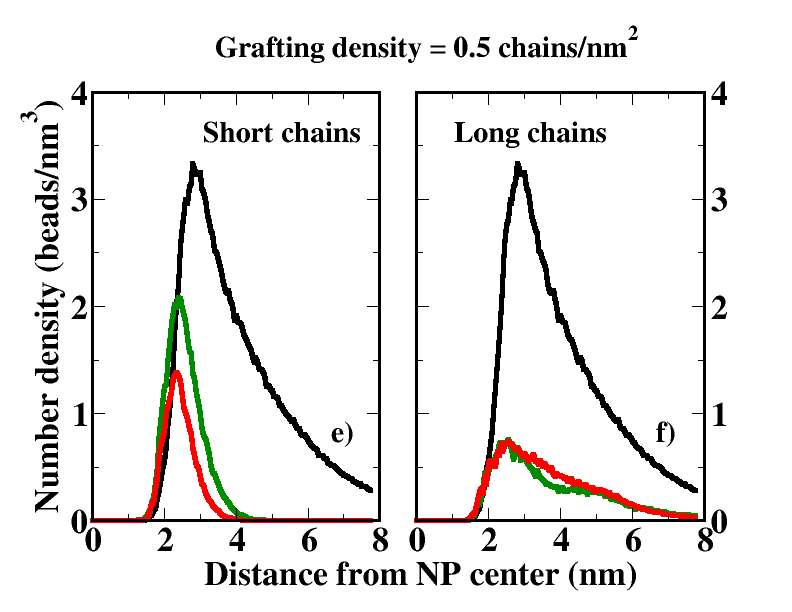}
\includegraphics[width=8.0cm,angle=0]{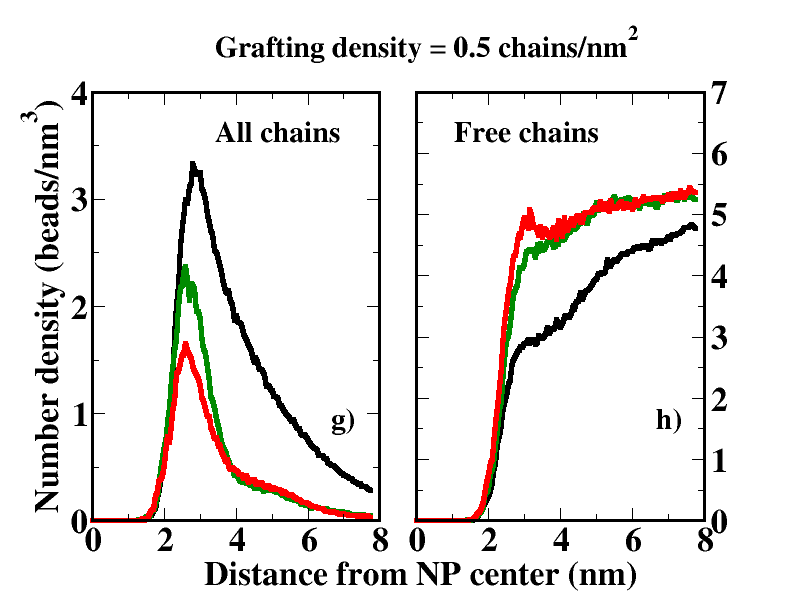} \\
\caption{Radial number density of monomers
belonging to grafted and free chains.
Two different grafting densities, namely
$\rho_g=0.1$ (top panels) and 0.5 (bottom panels) chains/nm$^2$ are reported. 
Short, long and all grafted chains are separately shown.
}
\label{fig:F1c}
\end{center}
\end{figure*}
%%%%%%%%%%%%%%%%%%%%%%%%%%%%%%%%%%%%%%%%%%%%%%%%%%%%%

We now focus on the effect of increasing $BDI_g$, leaving $BDI_f$
unchanged. 
In Fig.~\ref{fig:F1c} we analyze the radial number density of monomers 
for $\rho_g=0.1$ (top panels) and 0.5 (bottom panels) chains/nm$^2$
upon increasing $BDI_g$ from 1 to 3.25.
For $\rho_g=0.1$ chains/nm$^2$ the monomer number densities of short 
(panel (a)) and
long (panel (b)) chains oppositely behave, 
%upon increasing $BDI_g$, 
similarly to the trend observed upon increasing $BDI_f$. This
circumstance suggests that for $\rho_g=0.1$ chains/nm$^2$ the dependence of 
the two-body PMF on $BDI_g$ is expected to be similar to that observed on
$BDI_f$. In addition, also in this case the global behavior of all grafted
chains (panel (c)) and free chains (panel (d)) is not significantly
influenced by the bidispersity.

For $\rho_g=0.5$ chains/nm$^2$, the behavior of the monomer 
number density 
for short chains (panel (e)) is very similar
to that observed for 
$\rho_g=0.1$ chains/nm$^2$, 
whereas for long chains
(panel (f)) the profiles for $BDI_g=2.36$ and $BDI_g=3.25$ are practically 
coincident. Hence, in this case the decrease of the density profile of short
chains is not offset by the increase of the corresponding profile of long
chains and the distribution of all grafted chains (panel (g)) 
shows a dependence on $BDI_g$, decreasing when the latter increases. The
reverse behavior is found for free chains (panel (h)).  

\begin{figure*}[t!]
\begin{center}
\begin{tabular}{cc}
\includegraphics[width=8.0cm,angle=0]{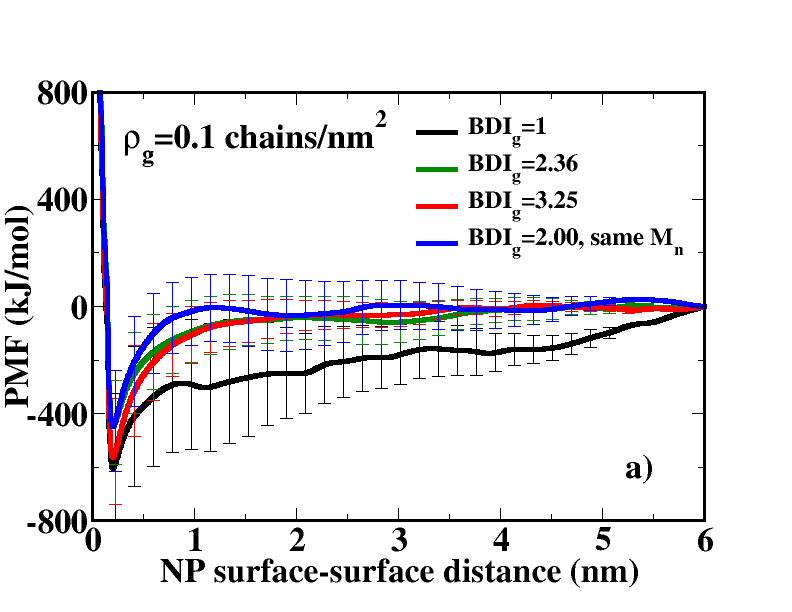}
\includegraphics[width=8.0cm,angle=0]{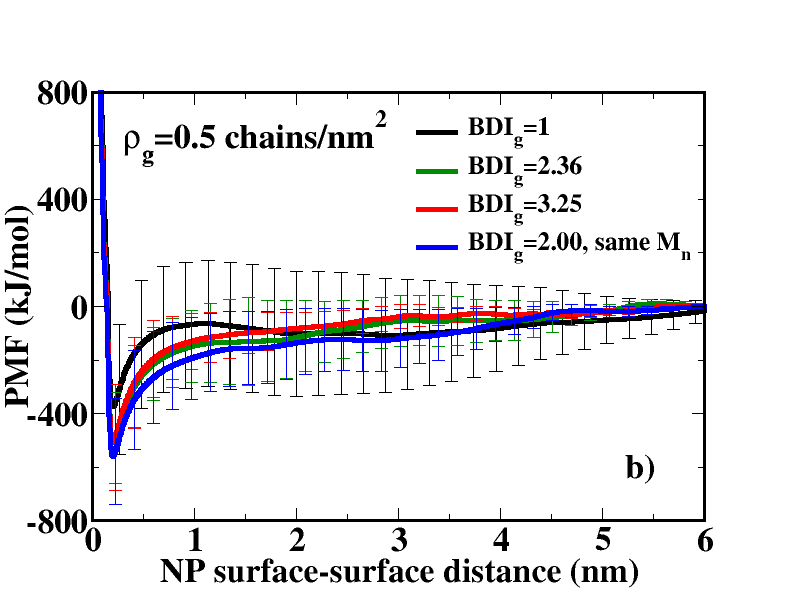}
\end{tabular}
\caption{Effect of increasing the bidispersity index of grafted chains 
on the potential of mean force between two grafted nanoparticles 
with grafting density $\rho=0.1$ (a) and 0.5 (b) 
chains/nm$^2$. In all cases the free chains are monodisperse.}
\label{fig:F3}
\end{center}
\end{figure*}

The two-body PMF as a function of $BDI_g$
is shown in Fig.~\ref{fig:F3}.
It is worth pointing out that, as anticipated in the 
{\bf System preparation} subsection, 
for bidisperse grafted chains a dependence on 
$M_n$ can be expected; 
hence, we have also calculated the 
two-body PMF for $\rho_g=0.1$ and 0.5 chains/nm$^2$ keeping 
this quantity constant.
For $\rho_g=0.1$ chains/nm$^2$ (panel (a)),
upon increasing $BDI_g$ a non-monotonic increase of the 
repulsion between the two NPs is found.  
This repulsion  
becomes stronger for $BDI_g=2.36$ than for $BDI_g=3.25$.
A comparison with the corresponding monomer number density 
(see panels (a) and (b) of 
Fig.~\ref{fig:F1c}) suggests that 
the PMF is more repulsive if the arrangement of long grafted chains around
the NP surface worsens (in comparison with the monodisperse case) and vice
versa.
The non-monotonic 
behavior of the PMF upon increasing $BDI_g$ also suggests
that in this
case two different effects (attraction due to the  
decrease of the length of the grafted chains,
and repulsion due to the increase of the wettability) compete, 
giving rise to a PMF slightly more repulsive than in the monodisperse case, but 
still exhibiting an attractive well for short interparticle separations. 
These effects can be ascribed almost entirely to the increase
of $BDI_g$; indeed, the behavior of the 
PMF calculated for $BDI_g=2.00$, 
with $M_n$ kept constant, is very similar to those already
discussed.
%%%%%%%%%%%%%%%%%%%%%%%%%%%%%%%%%%%%%%%%%%%%%%%%%%%%%
%%%%%%%%%%%%%%%%%%%%%%%%%%%%%%%%%%%%%%%%%%%%%%%%%%%%%
It is worth comparing the results obtained so far with an experimental
study devoted to investigate
the influence of the 
polydispersity of the grafted chains on the behavior of 
silica-PS nanocomposites in the case of low grafting 
densities~\cite{Natarajan:13}. 
It is known that for monodisperse distributions of free and grafted 
PS chains and low
grafting densities, NPs can self-assemble in a large variety of different
super-structures, including strings, connected sheets or small 
clusters~\cite{Kumar:13}.
Upon comparing their results with
those already existing for monodisperse polymer chains~\cite{Akcora:09}
the authors observe that for bidisperse  grafted chains the NPs
appear well dispersed in a more extended region of the phase diagram.
Hence, they suggest that a good recipe for increasing the repulsion between
grafted NPs is to use a bidisperse distribution of grafted PS chains,
as in the present case.
Our results are in agreement with these experimental findings. 
Interestingly, a similar behavior was observed also in an 
earlier Monte Carlo study of the bidispersity in the 
length of the grafted chains and its
effects on the PMF between NPs in homopolymer melts~\cite{Nair:11}.
Indeed in that work the authors found that the repulsion between NPs increases
with $BDI_g$. A subsequent theoretical-simulation 
study~\cite{Martin:13a} has shown that NPs with bidisperse distributions where 
the number of long grafted chains is less than the number of short grafted 
chains ({\it i.e.} the same prescription adopted in the present work) 
is more efficient in stabilizing the nanocomposites against the 
aggregation in comparison to both unimodal 
and polydisperse distributions. 
Finally, by looking at the behavior of the two-body PMF for 
$BDI_f=2.36$ (Fig.~\ref{fig:F2}a) and for $BDI_g=2.36$ and 3.25 
(Fig.~\ref{fig:F3}a) we also observe the existence of a short-range
attraction followed by a repulsion for intermediate NP-NP 
separation:
as described in Ref.~\cite{Munao:18a}, this picture is compatible
with the onset of self-assembled structures in the system, as
also experimentally detected~\cite{Akcora:09,Akcora:10,Kumar:13}.

The effect of increasing $BDI_g$ on the PMF between two
grafted NPs for $\rho_g=0.5$ chains/nm$^2$ is presented in
Fig.~\ref{fig:F3}b. Unlike all cases documented so far,
the increase of $BDI_g$ generates an increase of the attraction between the 
NPs. 
Indeed, since now the grafting density is 
higher, the free chains hardly penetrate the grafted corona and the
wettability is low; on the other hand, the reduction of the grafted chain
length upon increasing $BDI_g$ is significant (see Tab.~\ref{tab:2PMF}) and
hence the attractive contribution to the PMF is dominant. 
As observed for $\rho_g=0.1$ chains/nm$^2$, also in this case
the differences with respect to the unimodal 
distribution can be attributed
to the increase of the bidispersity, rather than to the different values of
$M_n$.
Note that in all
cases the PMF is entirely attractive and no nanostuctured aggregate can be
expected. This is confirmed by experiments~\cite{Akcora:10} 
and simulation~\cite{Akcora:09} studies, where it
was found that complex aggregates are observed only for low grafting densities
(i.e. 0.1 chains/nm$^2$), whereas only phase separation or well dispersed
conditions can be expected for intermediate-to-high grafting densities.

Summarizing the pictures emerging from 
Fig.~\ref{fig:F2}a and~\ref{fig:F3}a it appears that a good strategy for
obtaining a well dispersed state for $\rho_g=0.1$ chains/nm$^2$
is to significantly 
increase the bidispersity of free
chains, leaving $BDI_g$ unchanged. On the other hand, a moderate increase of
$BDI_f$ or $BDI_g$  may allow the system to have an 
onset of more complex structures, 
not observed in the monodisperse case. 
Further information on the role played by the wettability 
is 
gained by calculating the penetration depth of the free chains 
into the grafted corona: this quantity 
shows a noticeable correlation with the behavior of the two-body
PMF (see the {\bf S.I.}).

\subsection{Expected phase behavior}

The phase behavior of the nanocomposite can be further characterized by
calculating the second virial coefficient $B_2$ from the PMF; in particular,
it is known that positive values of $B_2$ indicate that
repulsive contributions are dominant,
whereas its negative values suggest a prevalence of attractive 
interactions~\cite{Hansennew}. This circumstance sets 
the second virial 
coefficient as a suitable tool for quickly predicting the changes in the
phase behavior of self-assembling systems~\cite{Munao:JPCM,Prestipino:17}.
The general definition of the
second virial coefficient for a potential without an angular dependence 
can be written as~\cite{McQuarrie:76}:
\begin{equation}\label{eq:b2}
B_2=2\pi \int_0^{\infty} (1-e^{-\beta U(r)}) r^2 dr \,,
\end{equation}
where $U(r)$ is the PMF, $\beta= 1/k_{\rm B} T$ and $r$ is the interparticle 
separation. In our case the interval of integration $[0,\infty]$ is replaced 
by $[d_{\rm min},d_{\rm max}]$ according to the definition of Eq.~\ref{eq:PMF}.
It may be worth pointing out that the calculation of $B_2$ can not discriminate
between a phase separation and the onset of aggregation, since in both cases
$B_2 < 0$. However the behavior of the PMF is different in these two
situations, since in the first case it 
is attractive in the whole range of
interparticle separations, whereas in the second case it is short-range
attractive and long-range repulsive, as largely documented in 
literature~\cite{Stradner,Attr_Rep,liu:11}.
%%%%%%%%%%%%%%%%%%%%%%%%%%%%%%%%%%%%%%%%%%%%%%%%%%%%%%
\begin{figure}[t!]
\begin{center}
\includegraphics[width=12.0cm,angle=0]{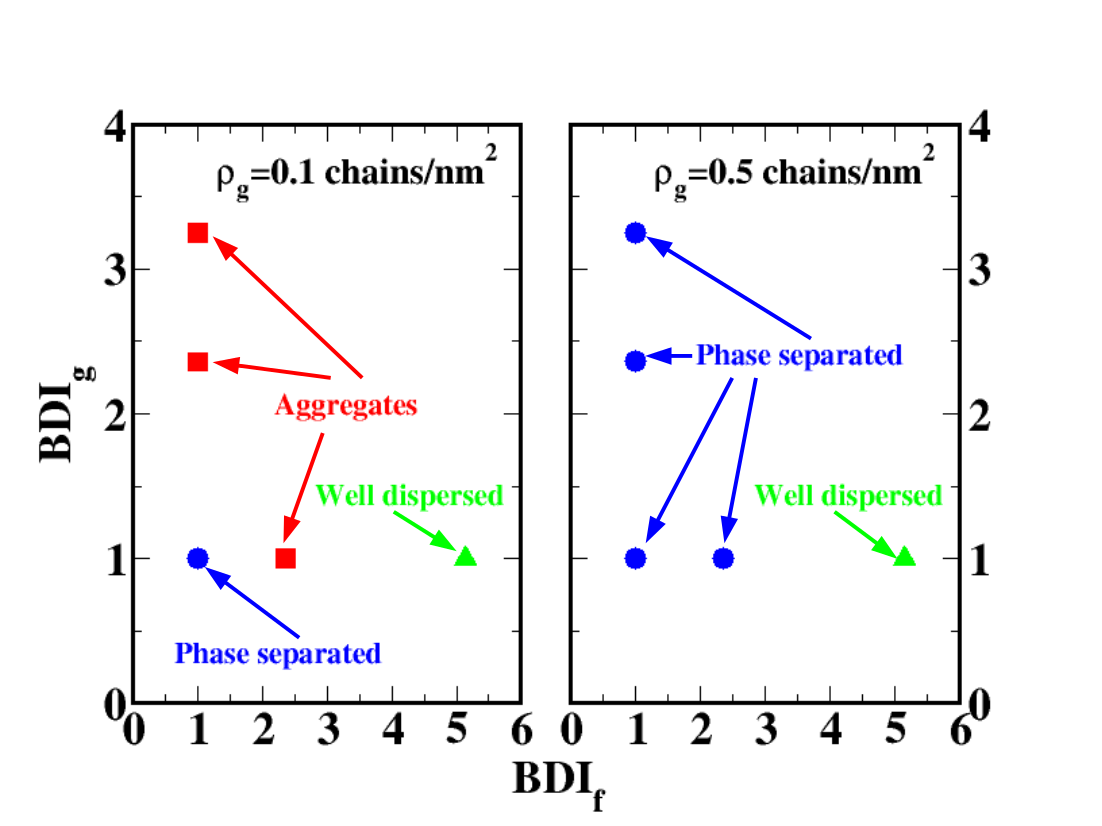}
\caption{Schematic phase behavior of grafted nanoparticles as a function of
bidispersity and grafting density. The diagram is drawn by considering the
behavior of the two-body PMF and the second virial coefficient.
Different colors and symbols identify phase separated ($B_2 < 0$),
aggregated ($B_2 < 0$) and well dispersed ($B_2 > 0$) conditions.}
\label{fig:F5a}
\end{center}
\end{figure}
%%%%%%%%%%%%%%%%%%%%%%%%%%%%%%%%%%%%%%%%%%%%%%%%%%%%%
The resulting, schematic, phase diagram is reported in Fig.~\ref{fig:F5a}:
it clearly emerges that for $\rho_g=0.1$ chains/nm$^2$, the increase of
$BDI_f$ or $BDI_g$ causes a corresponding increase of the repulsion, 
favouring the onset
of aggregates or well dispersed conditions that replace the phase separation
observed in the monodisperse case. On the other hand, for $\rho_g=0.5$ 
chains/nm$^2$ the phase separation still persists upon increasing 
$BDI_g$ and only a large $BDI_f$ leads to well-dispersed conditions.

These effects
are expected to be even more significant if multi-body contributions to the
PMF are taken into account. In order to ascertain the importance of these
contributions, we present in the next section results obtained by
calculating the three-body PMF, comparing them with the two-body case
and the experimental data.

\subsection{Three-body potential of mean force}

%%%%%%%%%%%%%%%%%%%%%%%%%%%%%%%%%%%%%%%%%%%%%%%%%%%%%%
\begin{table*}[t!]
\begin{center}
\caption{Nanocomposite systems investigated for the calculation of the
three-body potential of mean force.
The grafting density $\rho_g$ is in unit of chains/nm$^2$. 
The chains length is given in number of beads.
The box lengths 
are $L_x=L_y=L_z=22$ nm.
}\label{tab:3PMF}
\begin{tabular*}{1.05\textwidth}{@{\extracolsep{\fill}}ccccccccccccc}
\hline
\hline
& $\rho_g$ &  No$^{\circ}$ of & No$^{\circ}$ of
& Short   & Long  & No$^{\circ}$ of & No$^{\circ}$ of &
 Short & Long & $BDI_g$ & $BDI_f$ \\
&  &  short & long
&  grafted & grafted & short &  long &
 free & free &  &  \\
&  &  grafted & grafted
&  chains & chains & free &  free &
 chains & chains &  &  \\
&  &  chains & chains
&  length & length & chains &  chains &
 length & length &  &  \\
\hline
& 0   &  0 &  0 & 0 & 0 &  2796 & 0&  20 & 0 & 1 & 1 \\
& 0   &  0 &  0 & 0 & 0 &  2200 & 60&  20 & 200 & 1 & 2.36 \\
& 0.1   &  0 &  5 & 0 & 80 &  2775 & 0&  20 & 0 & 1 & 1 \\
& 0.1   &  0 &  5 & 0 & 80 &  2178 & 60&  20 & 200 & 1 & 2.36 \\
& 0.1   &  0 &  5 & 0 & 80 &  264 & 50&  20 & 1000 & 1 & 5.14 \\
& 0.1   &  4 &  1 & 10 & 80 &  2775 & 0&  20 & 0 & 2.36 & 1 \\
& 0.1   &  4 &  1 & 5 & 80 &  2775 & 0&  20 & 0 & 3.25 & 1 \\
& 0.1   &  4 &  1 & 80 & 1000 &  2595 & 0&  20 & 0 & 2.94 & 1 \\
& 0.5   &  0 &  25 & 0 & 80 &  2527 & 0&  20 & 0 & 1 & 1 \\
& 0.5   &  0 &  25 & 0 & 80 &  1987 & 54&  20 & 200 & 1 & 2.36 \\
& 0.5   &  20 &  5 & 10 & 80 &  2691 & 0&  20 & 0 & 2.36 & 1 \\
\hline
\end{tabular*}
\end{center}
\end{table*}
%%%%%%%%%%%%%%%%%%%%%%%%%%%%%%%%%%%%%%%%%%%%%%%%%%%%%%

\begin{figure}[t!]
\begin{center}
\includegraphics[width=8.0cm,angle=0]{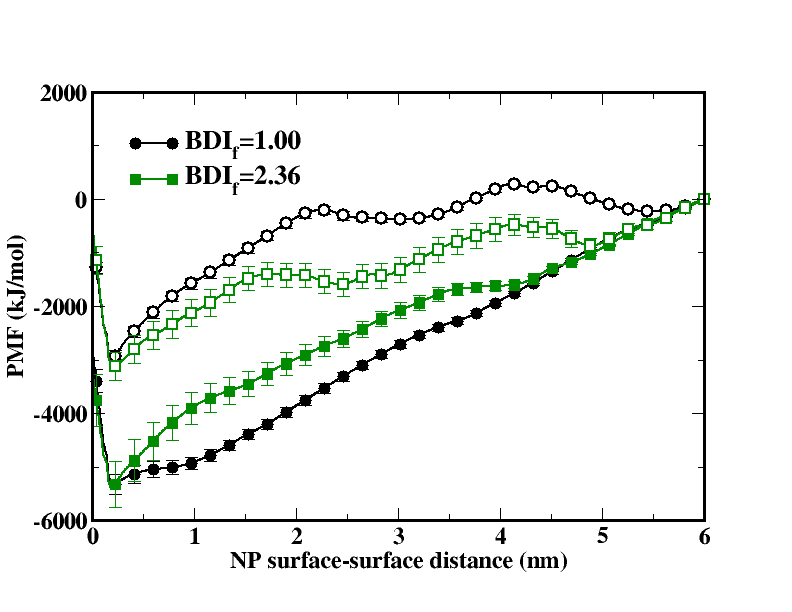}
\caption{Comparison between three-body (full symbols) and
two-body (open symbols) potential of mean force 
between ungrafted nanoparticles upon increasing the bidispersity index
of the free chains.
}
\label{fig:3bnak}
\end{center}
\end{figure}

The complete collection of systems 
investigated for the calculation of 
the three-body PMF is shown in 
Tab.~\ref{tab:3PMF}. The simulation box is now cubic in order to allow a proper
arrangement of the third NP to occur.
In Fig.~\ref{fig:3bnak} we compare three-body and two-body PMFs
between ungrafted NPs for $BDI_f=1.00$ and $BDI_f=2.36$. It emerges that
the three-body PMF is remarkably more attractive than the simpler two-body
interaction, due to the presence of the third NP in close contact with the
first two. This behavior holds regardless of the bidispersity and is
primarily due to the fact that a 
very low number of chains can be arranged between the
NPs, as explained also in Ref.~\cite{Munao:18a}. Also, the oscillations 
observed in the two-body PMF disappear in the three-body interaction:
this is due to the perturbation caused by the third NP to the previously
described mechanism of association, which gives rise to an attraction 
monotonically increasing upon decreasing the interparticle distance. 

%%%%%%%%%%%%%%%%%%%%%%%%%%%%%%%%%%%%%%%%%%%%%%%%%%%%%%
\begin{figure}[t!]
\begin{center}
\begin{tabular}{cc}
\includegraphics[width=8.0cm,angle=0]{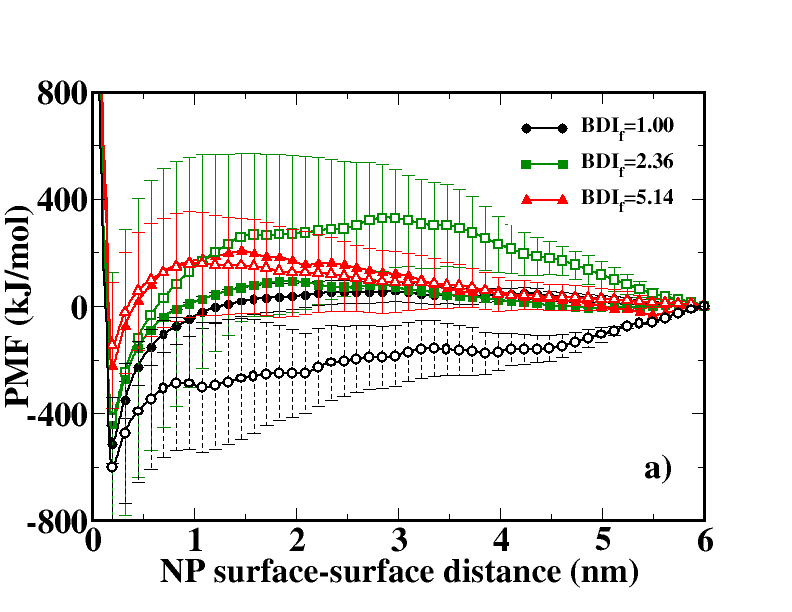}
\includegraphics[width=8.0cm,angle=0]{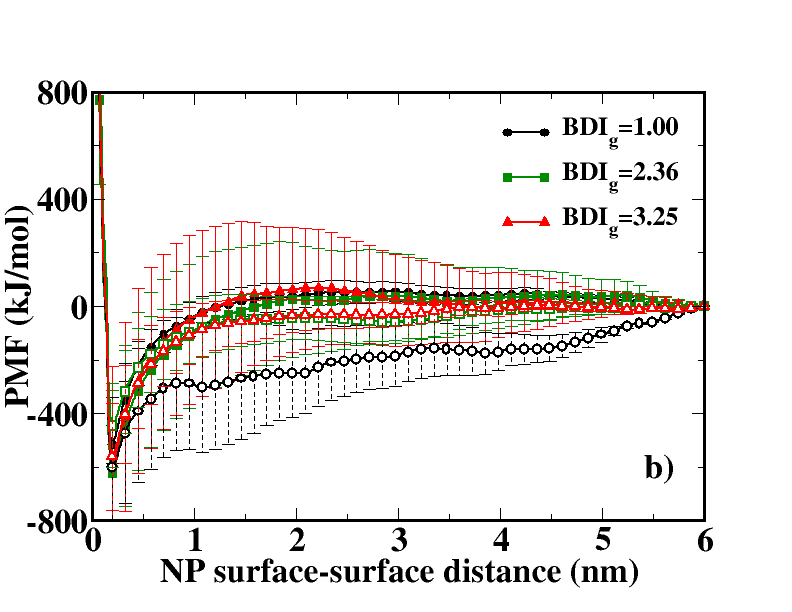}
\end{tabular}
\caption{Panel (a): comparison between three-body (full symbols) and
two-body (open symbols) potentials of mean force 
between grafted nanoparticles with grafting density 
$\rho_g$=0.1 chains/nm$^2$ upon increasing the bidispersity index of free 
chains. 
In all cases the grafted chains are monodisperse. 
Panel (b): same as panel (a) upon increasing the bidispersity 
index of the grafted chains, while the free chains are monodisperse.}
\label{fig:F6}
\end{center}
\end{figure}

%%%%%%%%%%%%%%%%%%%%%%%%%%%%%%%%%%%%%%%%%%%%%%%%%%%%%
Since the most interesting phase behaviors are observed for $\rho_g=0.1$
chains/nm$^2$, due to the possible appearance of self-assembled structures,
we investigate this specific grafting density in more detail. 
In Fig.~\ref{fig:F6}a we report the
three-body PMF for $\rho_g=0.1$ chains/nm$^2$ upon increasing $BDI_f$, 
comparing the results with the two-body PMF obtained in the same conditions
(see Fig.~\ref{fig:F2}a). In the monodisperse case, as also
observed in Ref.~\cite{Munao:18a}, the three-body PMF is remarkably more
repulsive than the two-body counterpart; 
for $BDI_f=2.36$ the trend appears reversed, 
whereas upon increasing $BDI_f$ to 5.14, two- and three-body 
PMFs similarly
behave. Hence, it appears that the multi-body effects and the increased
wettability of the NPs surface are combined in a non-trivial 
way, giving
rise to an effective interaction 
which is non-monotonically dependent on the bidispersity.
Furthermore, the differences between two- and three-body PMFs 
are progressively
less pronounced upon increasing $BDI_f$.
Interestingly, we finally note that the three-body cases studied are compatible
with systems where self-assembled structures arise, as experimentally
expected.

%%%%%%%%%%%%%%%%%%%%%%%%%%%%%%%%%%%%%%%%%%%%%%%%%%%%%%
%%%%%%%%%%%%%%%%%%%%%%%%%%%%%%%%%%%%%%%%%%%%%%%%%%%%%
The effect of increasing $BDI_g$ on the three-body PMF is analyzed in 
Fig.~\ref{fig:F6}b, along with the comparison with the two-body case.
As already observed upon increasing $BDI_f$, the effect of the bidispersity
on the three-body PMF is less evident than in the two-body counterpart: this
is indicated by the similarity between all the three-body PMF profiles.
On the other hand, in all cases the repulsion
increases when going from the two-body to the three-body interaction. 
This finding is in agreement with the results obtained for monodisperse
systems~\cite{Munao:18a} and also supports previously discussed 
experimental studies~\cite{Natarajan:13} where it was stated that
a good way for obtaining a good dispersion of NPs in the composite is to
increase $BDI_g$, provided that the grafting density is low. 

However, even if the three-body PMF is more repulsive than its two-body
counterpart, a good dispersion is not reached, since the
effective interaction still exhibits a deep attractive well. In this context
it is worth noting that in the calculation of 
the multi-body contributions to the
PMF, $M_n$ of the grafted chains is expected to play a 
more significant role than in the two-body case~\cite{Munao:18a}.
Hence, since for both $BDI_g$ investigated so far, $M_n$ has been 
decreased, we have performed further simulations for $BDI_g=2.94$ 
(see Tab.~\ref{tab:3PMF}) where $M_n$ has been increased. The results,
reported in the {\bf S.I.}, indicate that under these conditions the
three-body PMF is almost entirely repulsive and compatible
with a well dispersed condition.  

%%%%%%%%%%%%%%%%%%%%%%%%%%%%%%%%%%%%%%%%%%%%%%%%%%%%%%
\begin{figure}[t!]
\begin{center}
\includegraphics[width=8.0cm,angle=0]{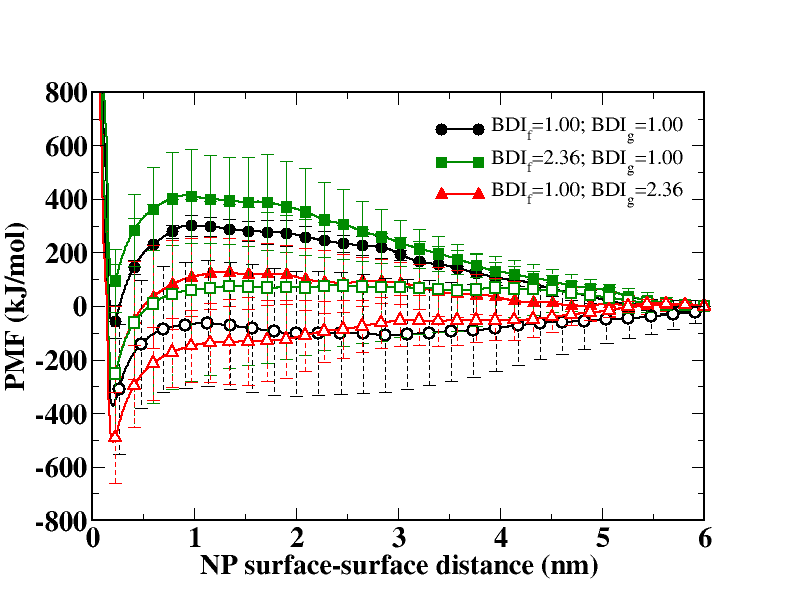}
\caption{Comparison between three-body (full symbols) and
two-body (open symbols) potentials of mean force 
between grafted nanoparticles with grafting density $\rho_g=0.5$ 
chains/nm$^2$ upon increasing the bidispersity index of free and grafted
chains.
}
\label{fig:3bgraf05}
\end{center}
\end{figure}
%%%%%%%%%%%%%%%%%%%%%%%%%%%%%%%%%%%%%%%%%%%%%%%%%%%%%
Finally, in Fig.~\ref{fig:3bgraf05} we analyze the effect 
of the bidispersity on the
three-body PMF for $\rho_g=0.5$ chains/nm$^2$. In this case the three-body PMF
is systematically more repulsive than the two-body interaction upon increasing
both $BDI_f$ and $BDI_g$. This is due to the repulsion between the grafted
chains, which increases if a third NP is placed close to the first two. 
Interestingly, the increase of $BDI_f$ promotes the repulsion, while the
increase of $BDI_g$ favours the attraction: this picture holds 
both for the
two-body and the three-body PMFs. 
This finding suggests that for this grafting
density the mechanisms governing the microscopic interactions are maintained
when going from a two-body to a three-body description of the PMF.

The refinement of the simpler two-body interactions and the 
qualitative agreement with the experimental results confirm that
the multi-body effects are not negligible in polymer nanocomposites, and
that a proper tuning of the bidispersity can be useful in view of
obtaining a desired structure.

\section{Conclusions}
In the present work we have studied the local structure and the effective
interactions in a polymer nanocomposite system constituted by silica
nanoparticles (NPs) in a polystyrene (PS) melt. NPs  
are assumed to be either 
bare or grafted with PS chains. The study has been performed by
applying the hybrid particle-field molecular dynamics approach, which has
allowed us to successfully simulate even systems with  
high molecular weight 
PS chains. The main focus of our investigation concerned the role
played by the bidispersity of the length of the polymer chains
in determining the final
phase behavior of the composite system. 
A comparison with previous atomistic~\cite{Ndoro:11,Ndoro:12,Eslami:13} 
and coarse-grained~\cite{Muller-Plathe:12} molecular dynamics 
simulation studies concerning monodisperse systems has shown a good agreement in
reproducing the local structure of polymer chains around the NPs.
By calculating the monomer number
densities of both free and grafted chains and the two- or multi-body 
potentials of mean force (PMF) we have ascertained 
that if the NPs are 
ungrafted, an increase of the bidispersity index ($BDI$) 
causes a stronger 
attraction between them. In the grafted cases, a crucial role is played
by the grafting density $\rho_g$: for $\rho_g=0.1$ chains/nm$^2$ 
the increase of $BDI$
of both free and grafted chains ($BDI_f$ and $BDI_g$, respectively) 
causes a rise of the repulsion between the NPs. 
Instead, for $\rho_g=0.5$ chains/nm$^2$ the repulsion increases only with
$BDI_f$ and diminishes if $BDI_g$ increases. 
It is worth noting that some trends observed in the present
study (such as the attraction between the ungrafted NPs and the increase
of the repulsion with the grafting density of grafted NPs) are not limited
to 
the specific case of silica-PS nanocomposites but are quite general.
The complex interplay between
bidispersity, grafting density and multi-body interaction gives rise to a 
rich phase behavior which qualitatively follows the available
experimental data on these systems. 
The capability of the hybrid particle-field approach 
to describe the microscopic interactions at the 
molecular level sets this method
as a valuable tool for investigations of the composite stability
under a large variety of conditions.

\section*{Supporting Information}
Calculation of the penetration depth of free chains inside the grafted
corona of a nanoparticle as a function of grafting density and bidispersity;
calculation of the three-body potential of mean force between 
nanoparticles grafted by polymer chains with high molecular weight.

\section*{Acknowledgements}
The computing resources and the related technical support used for this work
have been provided by CRESCO/ENEAGRID High Performance Computing infrastructure
and its staff~\cite{Cresco}. CRESCO/ENEAGRID High Performance Computing
infrastructure is funded by ENEA, the Italian National Agency for New
Technologies, Energy and Sustainable Economic Development and by Italian and
European research programs, see http://www.cresco.enea.it/english for
information. Gianmarco Muna\`o and Giuseppe Milano acknowledge financial 
support 
from the European Union Horizon 2020 Programme, under Grant Agreement
No. 760940. Andreas Kalogirou and Florian M{\"uller}-Plathe acknowledge 
support by the Deutsche Forschungsgemeinschaft through the Collaborative
Research Centre Transregio 146 ``Multiscale Simulation Methods for
Soft-Matter Systems.'.

%\bibliographystyle{biochem.bst}
%\bibliography{manuscript}
\providecommand{\latin}[1]{#1}
\makeatletter
\providecommand{\doi}
  {\begingroup\let\do\@makeother\dospecials
  \catcode`\{=1 \catcode`\}=2 \doi@aux}
\providecommand{\doi@aux}[1]{\endgroup\texttt{#1}}
\makeatother
\providecommand*\mcitethebibliography{\thebibliography}
\csname @ifundefined\endcsname{endmcitethebibliography}
  {\let\endmcitethebibliography\endthebibliography}{}

\end{document}